# On the formation of cores in accreting filaments and the impact of ambient environment on it

S. V. Anathpindika [1, 2*] and Di Francesco, J [3]
[1]Indian Institute of Technology, Kharagpur, West Bengal, India
[2]University Observatory München, Schneirstrasse 1, 81679, München, Germany
[3]National Research Council of Canada, Herzberg, Astronomy & Astrophysics Research Centre, 5071 West Saanich Road, Victoria (BC), Canada V9E 2E7

**Abstract**
Results from some recent numerical works, including ours, lend credence to the thesis that the ambient environment, i.e., the magnitude of external pressure, affects the star-forming ability of clouds and filaments. In continuation with our series of papers on this subject, we explore this thesis further by developing new hydrodynamic simulations of accreting filaments confined by external pressures in the range $10^{4-7}$ $K$ $cm^{-3}$. Our principal findings are - **(i)** irrespective of linemass, filament-fragmentation generally yields spheroidal cores. The initially sub-critical filaments in low to intermediate external pressure environments form broad cores suggesting that weakly self-gravitating filaments must fragment via the *collect - and- collapse* mode to form broad cores. Transcritical filaments, by contrast, become susceptible to the Jeans-type instability and form pinched cores; **(ii)** the ambient environment bears upon the physical properties of filaments including their $FWHM_{fil}$. Only the filaments initially suffused with subsonic turbulence in Solar-Neighbourhood-like environments, however, have $FWHM_{fil}\sim$ 0.1 $pc$. In high pressure environs such filaments not only have much smaller widths, but also become severely eviscerated. On the contrary, filaments suffused with initially supersonic turbulence are typically broader; **(iii)** the quasi-oscillatory nature of velocity gradients must be ubiquitous along filament lengths and its magnitude generally increases with increasing pressure. The periodicity of the velocity gradients approximately matches the fragmentation lengthscale of filaments; **(iv)** oscillatory features of the radial component of the velocity gradient are a unreliable proxy for detecting signatures of accretion onto filaments; and **(v)** filaments at either extreme of external pressure are inefficient at cycling gas into the dense phase which could reconcile the corresponding inefficiency of star-formation in such environments.

**Keywords:** ISM : Clouds Physical data and Processes : gravitation, hydrodynamics Stars : formation

# 1 INTRODUCTION

Bulk of the dense mass in molecular clouds (i.e., gas with visual extinction, $A_v \gtrsim 7$), appears filamentary with prestellar cores typically aligned along filament lengths projected on the plane of sky (e.g., Könyves *et al.* 2015, 2020). So the ubiquity of filaments in the interstellar medium (ISM) and their importance in the formation of stars is hardly in doubt. *Herschel* offered a wide dynamic range of observations in various far-infrared bands which revealed in unprecedented detail the filamentary organisation of gas in molecular clouds (André *et al.* 2010, Arzoumanian *et al.* 2011, Men'schikov *et al.* 2010). These observations also revealed that filaments seldom occur in isolated environments, and certainly do not resemble the static idealised cylinder often assumed in analytic models (e.g., Stodólkiewicz 1963, Ostriker 1964, Nagasawa 1987). On the contrary, accretion from the ambient environment appears to play a key role in determining the vital characteristics of filaments (e.g., Schneider *et al.* 2010, Kirk *et al.* 2013, Beuther *et al.* 2015, Gong *et al.* 2021).

Detailed surveys of the filaments in clouds of the Solar-Neighbourhood (i.e., within a radius of $\sim$500 $pc$), and those farther away suggest that filaments exhibit a wide range of physical properties. For example, the *Herschel* Gould Belt Survey showed that filaments in these clouds have a characteristic width (i.e., the $FWHM_{fil}$) of $\sim$0.1 $pc$ (e.g., André *et al.* 2010, Arzoumanian *et al.* 2011, 2019), with linemasses in the range 5 - 17 $M_{\odot}$ $pc^{-1}$, peak densities of $\sim$ 3 - 9$\times10^{21}$ $cm^{-2}$ and average dust-based temperatures between 14 - 16 $K$ (e.g., Arzoumanian *et al.* 2019, Orkisz *et al.* 2019, Suri *et al.* 2019). Moreover, observations with dense gas

*E-mail: sumed_k@yahoo.co.in (SVA)

tracers like $H^{13}CO^{+}$ have enabled detection of thinner constituent fibres within filaments (e.g., Hacar *et al.* 2018).

In contrast, Galactic surveys extending beyond the Solar-Neighbourhood, like the HiGAL survey for example, revealed longer filaments expanding over a few parsecs with much higher linemasses that typically range between several tens of $M_{\odot}$ $pc^{-1}$ to a few hundred $M_{\odot}$ $pc^{-1}$. These filaments are also generally much warmer than the typical *Herschel* Solar-Neighbourhood filaments (e.g., Schisano *et al.* 2020). Similarly, a number of IRDCs[1] also appear elongated and filament-like on the plane of sky. Many IRDCs are located at distances of a few *kpc* and typically have column densities upward of a few times $10^{21}$ $cm^{-2}$, and linemasses $\gtrsim 10^2$ $M_{\odot}$ $pc^{-1}$ (Peretto *et al.* 2014, Henshaw *et al.* 2014, Rathborne *et al.* 2016). Detailed analyses of their column density structures show further that several indeed have filamentary or hub-filamentary morphologies (e.g., Busquet *et al.* 2016, Henshaw *et al.* 2017, and the review by Hacar *et al.* 2022). Detailed analyses of the cloud G0.253+0.016, also known as the *Brick*, located in the Galactic CMZ revealed filaments with $FWHM_{fil} \sim$ 0.17±0.08 *pc* (Federrath *et al.* 2016).

A number of recent numerical Simulations explore filament formation at different spatial scales (e.g., Hennebelle 2013, Federrath *et al.* 2016, Chen *et al.* 2016, Inoue *et al.* 2018, Abe *et al.* 2021, Federrath *et al.* 2021). These simulations and others such as those by Smith *et al.* (2014) and Moeckel & Burkert (2015) show that filaments are a byproduct of the interaction between gas flows in a turbulent environment - the so -called paradigm of *turbulence-driven filament - formation (TDFF)*. In this paradigm, filaments and the cores within them often appear to form almost simultaneously (e.g., Gong & Ostriker 2011, Goméz & Vázquez-Semadeni 2014). Numerical Simulations like those by Federrath (2016), for example, for typical conditions in the Galactic arms show filament width in the range 0.05 - 0.15 *pc* irrespective of the star-formation history. Interestingly, these widths are also consistent with the sonic scale, i.e., the lengthscale on which turbulence transitions from the supersonic to the subsonic regime (see also, Federrath *et al.* 2021).

In other works about evolution of individual filaments Gritschneder *et al.* (2017), for example, showed that a small density perturbation on an initially sub-critical filament amplifies over time and eventually fragments its natal filament. In this genre of simulations individual filaments fragment due to the growth of local density perturbations along their length - the so called *geometrical fragmentation of filaments*. Heigl *et al.* (2018), similarly, showed that an initially sub-critical filament becomes susceptible to the *sausage-type* instability and forms a broad core, i.e., a core that is bigger than the width of its natal filament. An initially supercritical filament, on the other hand, becomes susceptible to the *Jeans-type compressional instability* to form pinched cores, i.e., cores of size smaller than the width of its natal filament. These observations are consistent with the analytic findings by Pon *et al.* (2011) who showed that local perturbations along the axis of a filament amplify faster than its global contraction timescale.

Prior to this paper we numerically explored the impact of ambient external pressure ($P_{ext}$) on filament evolution (i.e., Anathpindika & Di Francesco 2020, 2021 - hereafter referred to as Papers I and II, respectively). Those papers were motivated by the suggestions that ambient environmental conditions some how affect physical properties of gas, including its ability to form stars. This seems to be true in our own Galactic disk (e.g., Rathborne *et al.* 2014b, Rice *et al.* 2016), as well as in the disks of some other nearby galaxies (e.g., Hughes *et al.* 2010, Hughes *et al.* 2013, Heyer & Dame 2015). Arguably our thesis about the ambient environment is limited by the assumption that it is equivalent to only the external pressure. In reality, however, the environment could also mean the extended diffuse gas of which an observed filament is just a part. Nevertheless, in this series of papers we are exploring the thesis that ambient environment must affect the evolution of individual filaments just as it also affects evolution of clouds (e.g., Anathpindika *et al.* 2018).

While filaments discussed in Paper I were accreting, those in Paper II were non-accreting. Naturally, $P_{ext}$ in Paper I was a combination of the thermal pressure of the gas being accreted by the filament and the ram-pressure due to its in-flow velocity. So $P_{ext}$ there was equivalently characterised by the Mach number of the inflowing gas (typically between 0.8-24), and the ratio of the temperature of the inflowing gas to the initial temperature of the gas within the filament (typically between 10-15). In Paper II, however, $P_{ext}$ was purely thermal and was determined only by the temperature of the medium confining the filament. Simulations discussed therein were generated for different choices of the ratio of the temperature of the external medium to the initial temperature of the gas within the filament (typically between 4-30).

Our principal findings from Papers I & II are : **(i)** The ambient pressure does in fact bear upon the morphology of filament fragmentation and the cores spawned by it; **(ii)** Sub-critical filaments in low-pressure environments ($P_{ext}/k_B \lesssim 10^4$ $K$ $cm^{-3}$) formed cores via the *collect-and-collapse* mode. Also, these cores were not only bigger than their natal filaments, but the core formation timescale is also comparable to or greater than the $e_{folding}$ timescale that is typically on the order of a few freefall times. In other words, core formation in this instance is relatively slow; **(iii)** At higher pressures more comparable to that in the Solar-Neighbourhood (i.e., between a few times $10^4$ $K$ $cm^{-3}$ to a few times $10^5$ $K$ $cm^{-3}$), however, an initially sub-critical filament contracts to acquire a centrally peaked density profile and forms cores via the *Jeans-type compressional instability*. These cores are pinched and form on a timescale comparable to or smaller than the freefall time; and **(iv)** At still higher pressures typically upward of $\sim 10^6$ $K$ $cm^{-3}$, however, the filaments rupture, i.e., they become severely eviscerated (i.e., losing substantial fraction of their mass) before eventually breaking up into disjointed fragments.

---

[1] Infrared Dark Clouds

### 1.1 Context of this paper vis - a - vis Papers I & II

Expanding on our work discussed in those papers, we now develop simulations of accreting filaments with a relatively wide range of accretion rates. Crucial questions that we explore here are : **(a)** Do accreting filaments also become unstable to the sausage-type instability like filaments that either accrete too little mass (Paper I), or none at all (Paper II)? **(b)** Are the observed fluctuations in velocity gradients ubiquitous and if so, are they indeed cospatial with fluctuations in the density field ? and **(c)** Is the observation in Paper I of a slight increase in the FWHM$_{fil}$ of filaments with increasing external pressure ($P_{ext}$) inconsistent with the analytic prediction of Fischera & Martin (2012) ?

The latter investigation of a correlation between the filament width and the external pressure is necessary because the apparent universality of filament width has been called into question by some observational findings. For example, larger filament widths, typically between $\sim$0.26 $pc$ and $\sim$0.34 $pc$, were reported in more distant ridges (e.g., Hennemann *et al.* 2012), and in the Galactic plane filaments (e.g., Schisano *et al.* 2014) using *Herschel*. Furthermore, Panopoulou *et al.* (2014) also observed filaments with larger FWHM$_{fil}$s in $^{13}$CO emission. Thinner elongated structures with FWHM$_{fil}$s$\lesssim$ 0.05 $pc$, on the other hand, have been identified among sub-filaments and fibres using interferometric observations of dense molecular tracers (e.g., Fernández-López *et al.* 2014, Hacar *et al.* 2018, Dhabal *et al.* 2018).

Indeed, Panopoulou *et al.* (2022) argued that filaments farther away must appear wider due to their poorer resolution. This suggestion no doubt contradicts much of the observational evidence in contemporary literature which shows that ambient environment, i.e., external pressure bears upon the physical properties of the density structure in the ISM. Interestingly, however, convergence tests by André *et al.* (2022) reinforce the conclusion by Arzoumanian *et al.* (2011, 2019) about the existence of a physical lengthscale on the order of $\sim$0.1 $pc$, at least for the filaments in the Solar neighbourhood. Results by André *et al.* (2022) therefore underscore the need for a theoretical framework to reconcile this lengthscale. In view of the suggestion that filament widths roughly correspond to the sonic length (e.g., Federrath 2016), it is plausible that the underlying variations in the local sonic length, or equivalently, in the local Jeans length of the filaments (Anathpindika & Freundlich 2015) generate the observed variation in filament widths.

We also explore in this paper the variation of the fraction of putative star-forming gas as a function of external pressure to understand the impact of ambient environment on the efficiency of star formation. The paper is organised as follows - the numerical method and the initial conditions for the simulations are presented in §2. The results from our numerical Simulations are then presented and discussed in §3 and §4, respectively. We conclude in §5.

## 2 NUMERICAL METHOD AND SET UP

Numerical Simulations discussed in this work were developed using the SPH code SEREN (Hubber *et al.* 2011), the features of which were described in Papers I & II. As in Papers I & II, in this work also we model a section of a typical filament as a cylinder of gas having initially uniform density, the length and radius of which are, respectively, $L_{fil} = 1$ $pc$ and $r_{fil} = 0.2$ $pc$ for different choices of the initial linemass, $f_{cyl}$[2], listed in column 7 of Table 1. Besides these, the filament is also characterised by the initial gas temperature, $T_{gas}$, listed in column 10 of Table 1. With the critical linemass, $M_{l_{crit}}$, known at the temperature $T_{gas}$, the initial mass of the filament listed in column 4 of Table 1 can be readily calculated.

Note that these are only fiduciary choices of the respective physical parameters meant to represent an early phase during the evolutionary cycle of typical filaments in nearby clouds, and in the more dense elongated IRDCs. We prefer to commence our simulations with a uniform density configuration so that the impact of external pressure on the final density profile of model filaments can be qualitatively and quantitatively studied. Gas within this cylinder is assumed to have the usual molecular composition of the Solar-Neighbourhood. This cylindrical distribution of gas is allowed to accrete mass radially.

The set-up of the cylinder and the envelope of gas to be accreted by it is confined by a jacket of particles representing the intercloud medium (ICM), which is then placed in a periodic-box having dimensions (0.66,0.66,1.08) $pc$. We employ ordinary sink particles so that particles exceeding the minimum resolvable density ($\sim 10^{-15}$ g cm$^{-3}$) are replaced with sinks. This choice of density threshold is good enough to represent typical protostellar objects i.e., the adiabatic contractional phase of a core that spans a density range between $10^{-13}$ g cm$^{-3}$ and $10^{-18}$ g cm$^{-3}$ during which it becomes optically thick (e.g., Larson 1973; Low & Lynden-Bell 1976). We note, however, that collating statistical properties of sink particles is not our extant interest. The truncated nature of our model filaments is immaterial because the length or radius of a filament has no bearing on its evolutionary sequence. For it is the density which determines the evolutionary timescale and the fragmentation lengthscale of a filament (e.g., Nagasawa 1987).

---

[2] $f_{cyl} = \frac{M_l}{M_{l_{crit}}}$, is the fractional linemass and $M_{l_{crit}}$ is the critical linemass corresponding to the temperature $T_{gas}$.

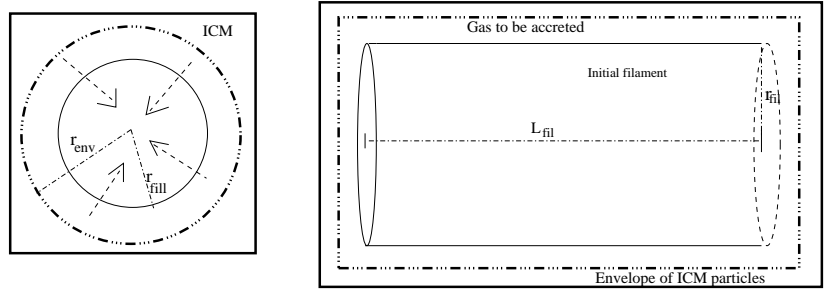

Figure 1.: Cartoon showing the schematic set-up of the test filament and the envelope of gas to be accreted by it. Shown on the left hand is a cross-section of this system; $r_{fil}$ and $r_{env}$ here denote the initial radius of the filament and the radius of the envelope of gas to be accreted, respectively.

Table 1: Physical parameters for simulations discussed below.

| Case No. | Pressure [$K\ cm^{-3}$] | | Mass[$M_\odot$] | | | | Temperature [$K$] | | | Velocity [$km\ s^{-1}$] | | $\dot{M}_{acc}$ |
|---|---|---|---|---|---|---|---|---|---|---|---|---|
| | $(P_{inf})^a$ | $(P_{ext})^b$ | $(M_{gas})^c$ | $(M_{acc})^d$ | $(M_{SPH})^e$ | $(f_{cyl})^f$ | $(T_{ICM})$ | $(T_{acc})$ | $(T_{gas})$ | $(V_{inf})$ | $(\sigma_g)$ | $(M_\odot\ Myr^{-1})^g$ |
| 1 | 5.7e3 | 2.6e4 | 1.65 | 10 | 2.4e-3 | 0.1 | 22.5 | 6 | 10 | 0.13 | 0.09 | 3.7 |
| 2 | 8.6e3 | 3.9e4 | 1.5 | 18 | 4.0e-3 | 0.1 | 20.25 | 5 | 9 | 0.12 | 0.09 | 6.2 |
| 3 | 1.2e4 | 4.1e4 | 3.6 | 10 | 2.8e-3 | 0.2 | 35.75 | 11 | 11 | 0.19 | 0.1 | 5.4 |
| 4 | 2.6e4 | 8.4e4 | 3.6 | 20 | 5.0e-3 | 0.2 | 24.75 | 11 | 11 | 0.19 | 0.1 | 10.9 |
| 5 | 4.5e4 | 2.0e5 | 82.0 | 1 | 1.0e-2 | 4.5 | 24.75 | 6 | 11 | 0.14 | 0.1 | 0.4 |
| 6 | 3.0e5 | 6.1e5 | 3.6 | 10 | 2.8e-3 | 0.2 | 453.75 | 220 | 11 | 0.87 | 0.1 | 24.9 |
| 7 | 5.11e5 | 1.27e6 | 3.6 | 20 | 5.0e-3 | 0.2 | 453.75 | 220 | 11 | 0.87 | 0.1 | 50.0 |
| 8 | 8.85e5 | 1.82e6 | 3.6 | 10 | 2.8e-3 | 0.2 | 1333.75 | 660 | 11 | 1.51 | 0.1 | 43.1 |
| 9 | 1.53e6 | 3.23e6 | 3.6 | 20 | 5.0e-3 | 0.2 | 1333.75 | 660 | 11 | 1.51 | 0.1 | 86.3 |
| 10 | 5.38e6 | 1.21e7 | 82.0 | 1 | 1.0e-2 | 4.5 | 1333.75 | 660 | 11 | 1.51 | 0.1 | 6.86 |
| 11 | 1.48e4 | 7.67e5 | 1.65 | 10 | 2.4e-3 | 0.2 | 572 | 11 | 11 | 0.19 | 1.36 | 8.9 |
| 12 | 1.7e4 | 8.97e5 | 1.5 | 18 | 4.0e-4 | 0.1 | 468 | 9 | 9 | 0.17 | 1.23 | 13.91 |
| 13 | 4.5e6 | 1.5e7 | 82 | 1 | 1.0e-2 | 4.5 | 1870 | 11 | 11 | 1.51 | 1.36 | 6.86 |

[a] Ram pressure of inflow.
[b] Pressure due to the externally confining intercloud medium (ICM).
[c] Filament mass.
[d] Accretional mass.
[e] Mass of individual SPH particle
[f] Initial linemass relative to the critical linemass at temperature $T_{gas}$ (see text).
[g] Accretion rate.

The simulations discussed in this work have an initial average smoothing length $h_{avg} = 0.01\ pc$ so that the spatial resolution is $0.02\ pc$. Following Eqn. (2) in Paper II and with the present choice of filament dimensions, the number of gas particles in the computational domain is $N_{gas} \sim 245000$, while that of particles representing the externally confining intercloud medium (ICM) is $N_{ICM} \sim 455000$. The mass of individual SPH particle, $M_{SPH}$, for each simulation has also been listed in Table 1. By comparison, the Jeans mass at the minimum achievable temperature (i.e., $\sim 5\ K$), and the highest resolvable density (i.e., $\sim 10^{-15}\ g\ cm^{-3}$) in these simulations is $\sim 0.01\ M_\odot$, meaning that the Jeans mass is well resolved and satisfies the criterion defined by Hubber et al. (2006) which is the SPH equivalent of the Truelove criterion of minimum resolution in adaptive mesh codes. Even in the worst case scenario, i.e., in Cases 5, 10 & 13 the minimum resolvable mass is comparable to the Jeans mass meaning that we are unlikely to see fictitious fragmentation. Simulations in this work were developed with periodic gravity. The cartoon in Fig. 1 shows a schematic representation of this set-up.

$M_{acc}$ is the mass of the gas available to be accreted by the filament which is the third free parameter in this set-up besides the initial temperature of the gas in the filament, $T_{gas}$, and the initial filament linemass, $f_{cyl}$. This is because we assume that the gas being accreted is transonic so that its Mach number, $\mathcal{M}_{inf} = 1$, and so its inflowing velocity, $V_{inf} = a_{inf}$, its sound-speed. The model filament is placed in a silo of gas to be accreted by it, the thickness of which is quantified in terms of the average SPH smoothing length and given in multiples of an integer parameter, $\eta$; here $\eta = 2$ across all simulations. Thus the rate of accretion, $\dot{M}_{acc} = \frac{M_{acc} \cdot V_{inf}}{(r_{fil} + \eta h_{avg})}$. The entire set-up is arranged such that there is initially approximate pressure equilibrium across the interface between the filament and the envelope of gas to be accreted. We assume no density contrast between gas in the filament and that to be accreted, or between the gas and the ICM. For any density contrast will naturally create an additional pressure gradient which will further affect filament evolution. We wish to avoid any such possibility.

The model filament here is initially superposed by sub-sonic turbulence (initial Mach number $\mathcal{M}^{int}$=0.5) with a power spectrum $\sim k^{-3}$. The choice of a relatively steep power-spectrum is consistent with that for the diffuse optically thin gas at intermediate Mach numbers (e.g., Stanimirović *et al.* 1999; Stanimirović & Lazarian 2001). The injected turbulence comprises of a natural mixture of the solenoidal and compressional modes. Thus the filament is supported against self-gravity by a combination of thermal pressure and pressure due to the turbulence. The expression for initial pressure balance is,

$$P_{int} = (P_{int}^{therm} + P_{int}^{turb}) \equiv P_{ext} = -(P_{inf} + P_{ext}^{therm}); \quad (1)$$

where $P_{int}$, $P_{ext}$, and $P_{inf}(\equiv \rho_{gas}V_{inf}^2)$ are, respectively, the internal pressure, the pressure due to the ICM, and the ram pressure due to the accretion of inflowing gas experienced by the filament; and $\rho_{gas}$ is the average initial density of the gas accreted by the filament. $P_{int}^{therm}$ and $P_{int}^{turb}$ are, respectively, the thermal component of internal pressure and the component of pressure due to internal turbulence. Finally, $P_{ext}^{therm}$ is the thermal component of pressure exerted by the inflowing gas. The -ve sign on the right hand side of Eqn. (1) indicates that this pressure is directed radially inward.

Rearranging Eqn. (1) yields,

$$T_{ICM} = \frac{(P_{int} + P_{inf} + P_{ext}^{therm})}{n_{ICM}}. \quad (2)$$

With $\mathcal{M}_{inf} = 1$ by choice, the initial gas temperature, $T_{gas}$, temperature of the gas being accreted, $T_{acc}$, the initial linemass, $f_{cyl}$, and the mass accreted, $M_{acc}$, are the only free parameters while the ICM temperature is calculated according to Eqn. (2). These physical quantities together contribute to the net external pressure denoted as $P_{ext}$ and listed in column 3 of Table 1. To test the convergence of results, we repeated this ensemble of simulations with five different random number seeds used to generate the initial turbulent velocity field that was initially overlaid on the filaments in the respective simulations. So we developed 65 simulations in all.

The motivation for this toy-model of accreting filaments has been discussed in the prequel to this paper. While the Mach number of the inflowing gas, $\mathcal{M}_{inf}$, could be crucial, the process of filament assembly is itself a subject of debate. It is unclear if turbulent flows directly assemble filaments (as in e.g., Federrath 2016), or if interacting turbulent gas-flows assemble sheets that subsequently fragment to form density structure. Recently, Rezaei Kh. *et al.* (2022), for example, suggested that the California MC is sheet-like. Such sheet-like clouds could appear filamentary when projected edge-on on the plane of the sky. Alternately, they could fragment to generate filaments, a mechanism completely different from filaments forming in a box of turbulent gas. So we do not introduce here another variable in the form of $\mathcal{M}_{inf}$. Listed in Table 1 are our choices of various parameters that allow us to span a range of external pressures between a few times $10^4$ $K$ $cm^{-3}$ to a few times $10^7$ $K$ $cm^{-3}$.

For context, the ambient pressure is the lowest in the farthest regions of the Galactic disk and it is upward of $\sim 10^7$ $K$ $cm^{-3}$ towards the Galactic Central Molecular Zone (CMZ). Observational works by (e.g., Goodman *et al.* 1993; Caselli *et al.* 2002; Pineda *et al.* 2010 & Hacar *et al.* 2013) show that relatively dense sub-structure, i.e., clumps & cores, in the Solar-Neighbourhood are thermally supported while turbulent non-thermal motions dominate in the transition region between the dense clumps/cores and the surrounding ICM. For our own MW galaxy the average mid-plane pressure is estimated to be $\frac{(P_{ext}/k_B)}{[K\ cm^{-3}]} \sim 3 \times 10^4$, and includes contributions in roughly equal measures from the diffuse thermal pressure, magnetic field pressure, non-thermal pressure and pressure due to cosmic rays (e.g., Boulares & Cox 1990; Slavin & Cox 1993; Oey & García - Seguera 2004). The pressure is of course higher in active star-forming regions (e.g., Henshaw *et al.* 2014). The co-existence of a warm tenuous ICM with the cold dense clumps/cores has led authors to conjecture that MCs themselves constitute a multi-phase medium (e.g., McKee 1995 and other references therein). The process of segregation of the respective phases in a purely hydrodynamic calculation was demonstrated in an earlier contribution (Anathpindika 2015).

Estimates of the average external pressure do not reflect the possible effects of turbulence induced variations in magnetic field strengths. Seta & Federrath (2022), for instance, show that turbulent dynamo, i.e., the generation/amplification of magnetic field in the interstellar medium as the gas is randomly stretched by a turbulent velocity field, can affect the strength of the magnetic field. Studying the impact of such localised enhancement of magnetic field on filament-evolution is, however, beyond the scope of this work as we assume an isotropic external pressure that is a combination of only the thermal and the ram-pressure of the inflowing gas.

For the more violent environment such as towards the Galactic CMZ, Federrath *et al.* (2016), for instance, derive 3D turbulent sonic Mach number of 11±3 for gas in the *Brick*. This Mach number, though comparable to that for typical molecular clouds in the Solar-Neighbourhood, corresponds to extreme pressure towards the CMZ. The assumed $T_{acc}$ in Table 1 (Cases 8 - 10) is consistent with the observationally inferred gas temperature towards the CMZ (e.g., Rodríguez-Fernández *et al.* 2001, Mills & Morris 2013, Ao *et al.* 2013, Rathborne *et al.* 2014a). Ao *et al.*, for example, confirmed warm dense gas at temperatures ~100 $K$ through observations of formaldehyde while Rodríguez-Fernández *et al.* reported hot dense gas (temperature ~ 400 - 600 $K$, and density $\lesssim 10^6$ $cm^{-3}$) in gas envelopes around CMZ clouds through observations of rotational transitions of $H_2$. Similarly, an extremely hot gas component (temperature ~700 $K$) has also been inferred towards SgrB2, one of the most massive and densest clouds in the CMZ that is known to be forming massive stars, through observations of absorption lines corresponding to higher $NH_3$ transitions (e.g., Ceccarelli et al. 2002, Wilson et al. 2006).

The values of external pressure $P_{ext}$ so obtained for simulations 1 - 5 are consistent with those inferred observationally for clouds in the outer regions of the Galactic disk and in the Solar-Neighbourhood (e.g., Murphy & Myers 1985, Maddalena et al. 1986, Bally 1987, Loren 1989, Tatematsu et al. 1993). The warm gas in Cases 6 & 7 is representative of the ambience in an active local star-forming region like the Orion MC-complex. Previous observations also show that typical rates of accretion by filaments vary between a few 10s to a few 100s $M_\odot$ $Myr^{-1}$ $pc^{-1}$ (e.g., Kirk et al. 2013, Palmeirim et al. 2013, Schisano et al. 2014, Bonne et al. 2020, Gong et al. 2021). So the accretion rates assumed here are consistent with such inferred values.

For the sake of convenience, we classify the simulations discussed in this work into three types on the basis of the mass accreted by the respective filaments : **Type I (Cases 1, 3, 6, & 8)**, where the initially sub-critical filaments were allowed to accrete gas but still remained sub-critical (i.e., $f_{cyl} < 1$); **Type II (Cases 2, 4, 7, & 9 )**, where the initially sub-critical filaments were allowed to accrete enough gas such that they became super-critical (i.e., $f_{cyl} > 1$); and **Type III (Cases 5 & 10)**, the initially supercritical filaments, that of course remain supercritical. For reference, Table 2 summarises the nomenclature adopted for discussion of the simulations in the rest of this paper. Observe that the various cases discussed in this work have been listed in Tables 1 & 2 in the order of increasing pressure for each type. We repeat one simulation for each filament type with supersonic initial turbulence, i.e., $\mathcal{M}^{int}=7$. These are listed as respectively, Cases 11, 12 & 13 in Tables 1 & 2.

Finally, given the universality of the paradigm of structure formation via interaction between turbulent flows, the physics of filament formation must essentially remain the same across environments. So the model of an accreting filament assumed in this work, though simplified, is robust.

## 2.1 Limitations

The numerical set-up described above and indeed, in the sequel to this paper, is simplified because with some notable exceptions like the *Musca* filament (e.g., Hacar et al. 2016), isolated filaments are seldom observed in the field. Hub-filamentary structure, i.e., converging network of filaments are often reported in typical star-forming clouds (e.g., Peretto et al. 2021; Kumar et al. 2022). We have, however, preferred the relatively simple set-up of a singular cylinder here because our primary interest is to study the process of filament-fragmentation and not how individual filaments and/or bunch of filaments are assembled in MCs. This latter objective is itself the subject of study of a subsequent paper.

While a number of studies demonstrate the importance of magnetic field in the evolution of the ISM (e.g.,

Table 2: Classification of simulations listed in Table 1 on the basis of $f_{cyl}$.

| Case No. | External pressure $(P_{ext})$[a] | Type[b] |
|---|---|---|
| 1 | 2.6e4 | I |
| 2 | 3.9e4 | II |
| 3 | 4.1e4 | I |
| 4 | 8.4e4 | II |
| 5 | 2.0e5 | III |
| 6 | 3.2e5 | I |
| 7 | 1.27e6 | II |
| 8 | 1.82e6 | I |
| 9 | 3.23e6 | II |
| 10 | 1.21e7 | III |
| 11 | 7.67e5 | I |
| 12 | 8.97e5 | II |
| 13 | 1.5e7 | III |

[a]Column 3 in Table 1
[b]Type I ($f_{cyl} < 1$ even after accreting gas), Type II ($f_{cyl} > 1$ after accreting gas), Type III ($f_{cyl} > 1$ before accretion)

Padoan & Nordlund 2011; Krumbolz & Federrath 2019), the set-up here is purely hydrodynamic. The general consensus of these works, and indeed in other literature, is that the presence of magnetic field dampens the rate of star-formation although the actual impact is fairly modest and that the presence of magnetic field typically reduces the star-formation rate by only a factor of 2-3. In his analytic work, Nagasawa (1987) further showed that the magnetic field slowed the growth of the gravitational instability in a self-gravitating cylinder without affecting the wavelengths of unstable modes.

Evidently, the absence of magnetic field in this work is therefore unlikely to significantly alter the final outcome (except perhaps the fragmentation timescale) of the simulations discussed herein. However, we appreciate that magnetic fields in a turbulent medium could enhance filament-formation, though probably reduce the number of clumps, cores & stars. Conversely, by stabilising MCs against self-gravity magnetic fields could possibly induce formation of more massive stars and thus enhance the impact of stellar feedback (Hennebelle & Inutsuka 2019), leading to a higher external pressure, but still within the range of external pressure explored in this work.

## 3 RESULTS

### 3.1 GENERAL EVOLUTIONARY FEATURES OF THE FILAMENTS

We have seen in our earlier works that filaments evolve through a series of radial contractions, i.e., circular contractions in the direction perpendicular to the filament axis, followed by fragmentation along the axial direction, irrespective of whether they are accreting (as in Paper I where filaments accreted so little gas that linemass did not vary significantly), or non-accreting (Paper II). As with the filaments in those works, filaments in this work are also allowed to cool dynamically so that the resulting loss of pressure equilibrium triggers radial oscillations. The temporal evolution of filaments in this work is briefly described in Appendix A and the rendered density images in Figs. A1 - A3 show the evolutionary sequence of the filaments in some of the simulations developed in this work.

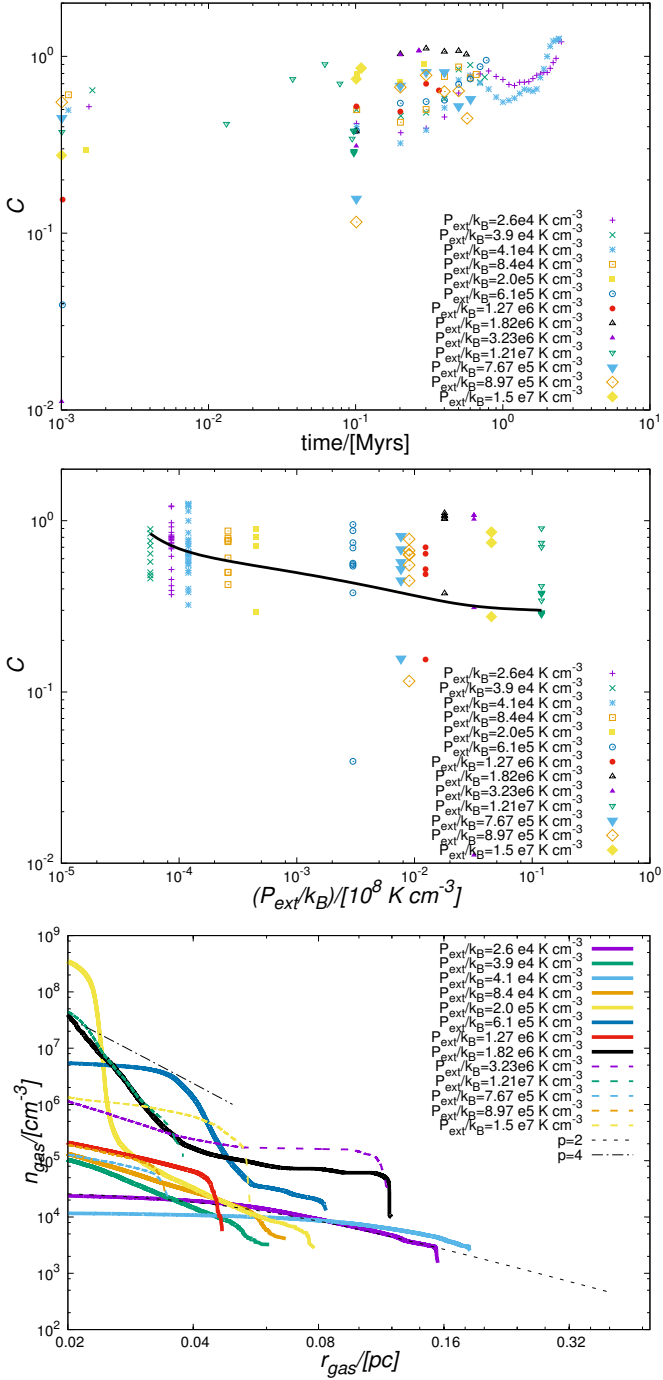

Figure 2.: *Top-panel:* Temporal evolution of the concentration parameter. *Central-panel:* The concentration parameter as a function of the external pressure. Data points on both these plots represent the value of $C$ at different epochs of filament evolution. The continuous line here represents the locus of mean values of $C$ for each simulation. *Lower-panel:* Radial density profiles of the filaments in respective simulations at the terminal epoch of their evolution.

The sub-critical **Type I** filaments evolve on rela-tively long timescales, i.e., on the order of a few free-fall times and comparable to the $e_{folding}$ timescale, while the super-critical **Type II & III** filaments evolve more rapidly, on timescales comparable to, or shorter than, the freefall time. As was seen in Papers I & II as well as in the semi-analytic work by Fiege & Pudritz (2002), for example, the filaments remain pressure truncated throughout the courses of their evolution. Like the latter authors, we calculate the concentration parameter, $C$, defined as $\log_{10}\left(\frac{r_{fil}(t)}{r_0(t)}\right)$, where $r_{fil}(t)$ is the outer radius of the filament at the epoch $t$, and the natural radial scale factor,

$$r_0(t) = \frac{a_{eff}(t)}{\sqrt{4\pi G \rho_c(t)}}, \qquad (3)$$

where $a_{eff}$ and $\rho_c$ are the effective sound speed and the central density calculated at an epoch $t$, respectively; $r_0(t)$ defines the effective core radius of the filament (Fiege & Pudritz 2002). Although analogous to the truncation parameter defined for the King models of globular clusters (e.g., Binney & Tremaine 1987), the tidal radius in its definition is replaced here with the outer filament radius, $r_{fil}$. The concentration parameter thus specifies the radius of truncation of a filament and is therefore a useful proxy for the filament width.

Figure 2 shows how $C$ evolves over time (top panel), and the external pressure (central panel), and how $n_{gas}$ varies with radius at the terminal epoch of the simulation (bottom panel). Plots in the top & the central panels show the mean value of $C$ across simulations developed with 5 different random number seeds. The density profiles being mutually similar across simulations with different random number seeds, we show the profiles from only one set. The temporal evolution of the filament radius can be easily inferred from that of $C$ as shown on the top panel of Fig. 2. Also evident from this panel is the fact that with the exception of the filaments in Cases 11-13, the concentration parameter decreases with increasing external pressure. This latter aspect, which essentially means that filaments (again with the exception of those in Cases 11-13), must become thinner with increasing external pressure, is more clearly visible from the plots in the central panel of Fig. 2. Naturally, the filaments in Cases 11-13 also have a bigger central radius, $r_{flat}$, as can be seen from the corresponding plots on the lower panel of Fig. 2. Evidently, the initially supersonic turbulence in the filaments in Cases 11-13 causes them to be puffed up. They thus have a bigger central radius. This observed variation in the inner radius is also likely to manifest itself in a similar trend for the filament width ($\text{FWHM}_{fil}$), a point that we will revisit later in §3.2.3.

Finally, the fact that these filaments evolve to have truncated density profiles irrespective of the external pressure is evident from the plot on the bottom panel. Note that these density profiles have been made after eliminating the SPH particles representing the ICM. The radial density distributions within filaments are largely

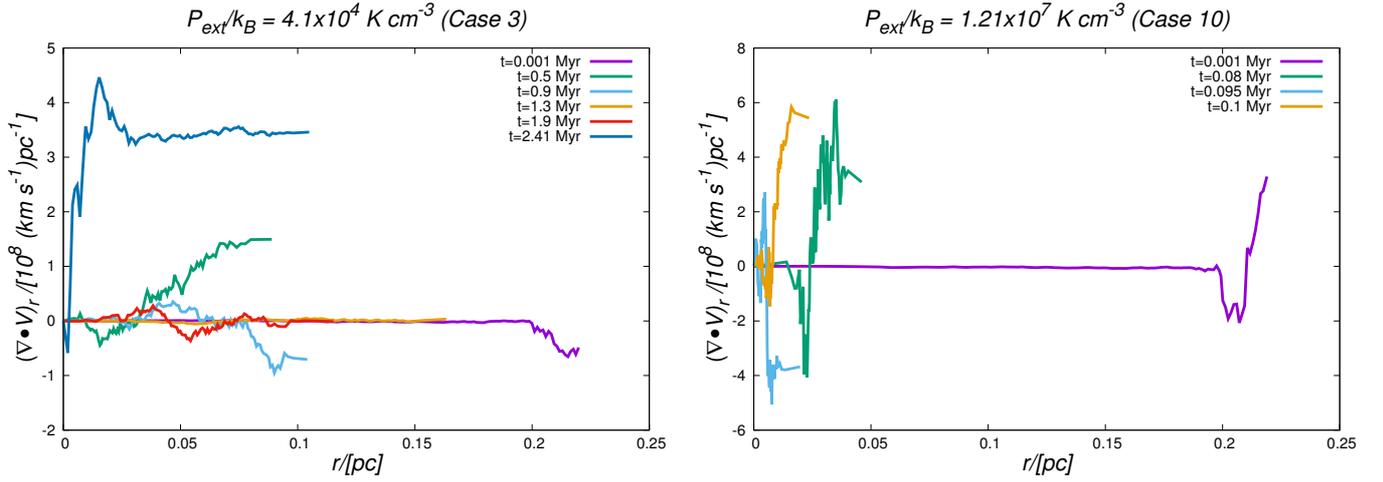

Figure 3.: Temporal variation of the radial component of the Divergence of the velocity field within the filament.

consistent with the Plummer profile given as,

$$\rho(r) = \frac{\rho_c}{[1 + (r/r_{flat})^2]^{\frac{p}{2}}}, \quad (4)$$

where $r_{flat}$ is the spatial extent of the inner flat portion of the density distribution, while the exponent $p$ varies between 2 and 4 for the magnitudes of external pressure considered here. Indeed, some of these density profiles also exhibit a distinct knee so that the profiles are shallower in the outer regions than closer to the centre, and the exponent $p$ also varies over the course of evolution of individual filaments. Interestingly, however, the exponent, $p$, becomes steeper with increasing external pressure. As previously noted, with the exception of the filaments in Cases 11-13, the central radius, $r_{flat}$, decreases with increasing external pressure, which is consistent with the trend between the concentration parameter, $C$, and the external pressure. In fact, $r_{flat}$ becomes vanishingly small for external pressures upward of $\gtrsim 10^6$ $K\ cm^{-3}$. Rendered density images in the Appendix (Figs. A1-A3) below illustrate the impact of ambient environment on the morphology of filament evolution. Collectively these images show, filaments, irrespective of their linemass, must become eviscerated in high pressure environs (as we shall also see in the following subsection). In low-pressure and Solar-like environs, however, weakly self-gravitating filaments must evolve relatively slowly and form broad cores via the *Collect and Collapse* mode.

The left- and right-hand panels of Fig. 3 show the temporal variation of the radial component of the divergence of the velocity field $(\nabla \cdot \mathbf{V})_r$ within the model filament in Cases 3 (**Type I** filament) and 10 (**Type III** filament), simulations representative of filament evolution in low- and high-pressure environments, respectively. $(\nabla \cdot \mathbf{V})_r$ is essentially the component of $(\nabla \cdot \mathbf{V})$ in the radial direction within the filament. Bear in mind that negative values of $(\nabla \cdot \mathbf{V})_r$ denote inward gas-motion, whereas positive values signify gas moving radially outward. For Case 3, the gradual inward contraction of the accreting filament over the course of its evolution is evident from the steadily decreasing $(\nabla \cdot \mathbf{V})_r$ towards the filament axis. Only towards the terminal epoch of the simulation when a core begins to collapse do we see a relatively strong inwardly directed velocity gradient. As with the plot in the bottom panel of Fig. 2, the plots shown on either panel of Fig. 3 are also from a simulation out of the ensemble developed with 5 different random number seeds for each case.

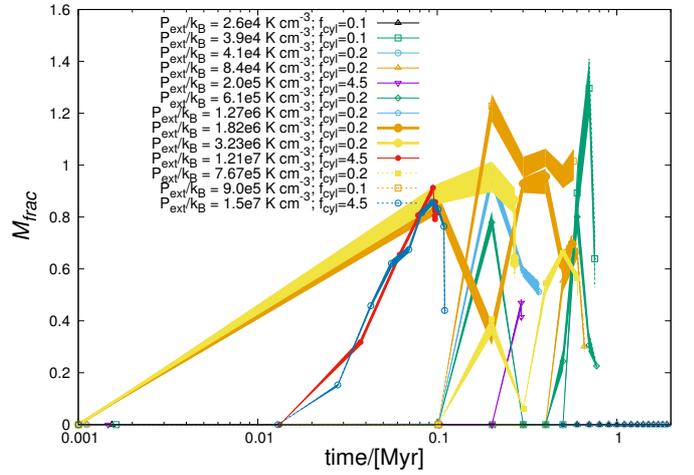

Figure 4.: Temporal variation of the fraction of mass having density $\gtrsim 10^{18}$ g cm$^{-3}$ in different simulations. Individual data points in this plot represent the mean value of $M_{frac}$ at each epoch. *See text for further explanation.*

For Case 10, by contrast, a radially inward travelling compressional disturbance is readily visible in the plot on the right-hand panel of Fig. 3. Note, however, that this disturbance is not a shock wave because the gas that is accreted is transonic ($\mathcal{M}_{inf}$=1). Indeed, the filament experiences ram-pressure due to this inflowing gas. The accreting filament contracts rapidly which causes pressure to build-up within it. The filament then relaxes, as can be seen from the temporal variation of $(\nabla \cdot \mathbf{V})_r$ in this plot. As the filament in either case evolves through a series of contractions and relaxations, some pockets of gas move inward while others move radially outward which is also significant from an observational perspective. This is because depending on the evolutionary stage of filament, pockets of gas at different locations within it could show signatures of radial expansion or contraction.

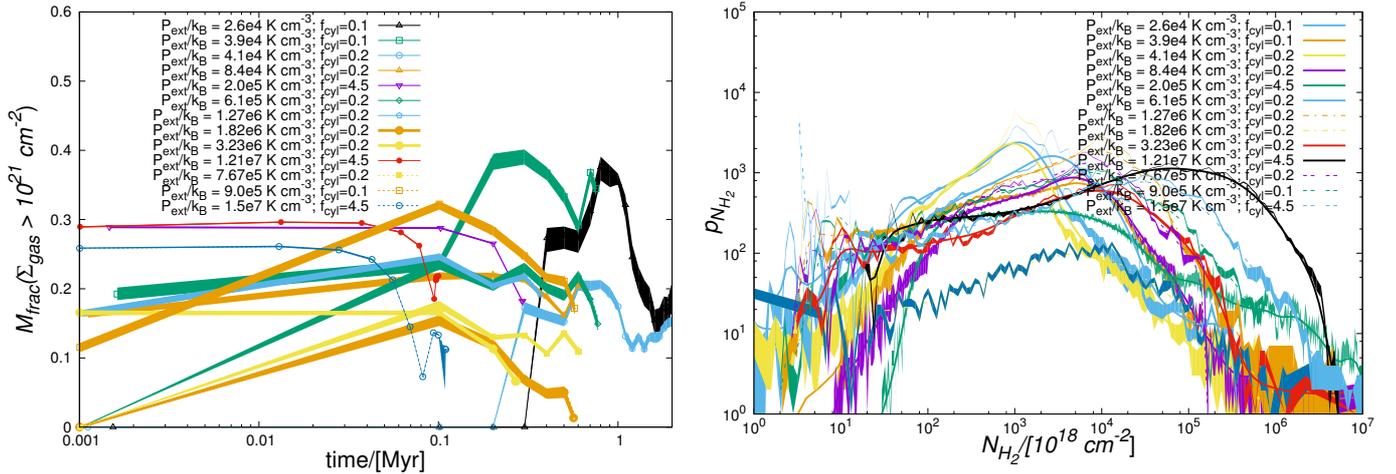

Figure 5.: *Left panel :* Temporal variation of the dense gas fraction in different simulations. *Right panel :* NPDF of molecular Hydrogen in filaments at their respective terminal epochs. As in Fig. 4 the lightly shaded region about each characteristic in either plot represents the variation of the respective quantities over the realisations developed with 5 random number seeds for different choices of the external pressure.

In the current set of simulations, the inwardly travelling compressional disturbance sets the filament boundary, or equivalently the filament width, unlike in the recent study by Priestley & Whitworth (2021), for instance, where the filament boundary is set by the location of the accretion shock. A decrease in filament width with increasing external pressure is clearly evident from Fig. 3 which also re-emphasises the observation made earlier regarding the central panel of Fig. 2 that shows the variation of the concentration parameter, $C$, as a function of the external pressure. This inference about possibly thinner filaments at higher external pressures, at least for filaments suffused with subsonic turbulence, though consistent with analytic predictions (e.g., Fischera & Martin 2012), seems at first to be inconsistent with the results we reported in Papers I & II. This inconsistency, as previously noted, is largely because the filament linemasses in those papers barely varied over a factor of a few in the course of their evolution. Naturally then, the self-gravity of those filaments did not significantly change, as they do here.

### 3.2 SOME TYPICAL PHYSICAL PROPERTIES

#### 3.2.1 DENSE GAS FRACTION

The dense gas fraction is a useful proxy to estimate the efficiency of star formation in molecular clouds (see, e.g., Lada *et al.* 2009). Here we invoke the same definitions used in an earlier work (i.e., Anathpindika *et al.* 2017) to quantify the fraction of gas in a filament cycled into the dense phase. Recall that we had then used two physical parameters, viz., the fraction of gas having volume density upward of typically $\sim 10^{18}$ $g$ $cm^{-3}$ and that having column density upward of $\sim 10^{21}$ $cm^{-2}$ to quantify the fractions of putative star-forming gas. We note that the column density is calculated by taking a projection of these filament in the plane orthogonal to the filament-axis.

Following the definitions of dense gas, Fig. 4 shows the temporal variation of dense gas fraction based on volume density, $M_{frac}$, in our simulations. As with the plots on the top & the central panels of Fig. 2, individual data points on this plot represent the mean of $M_{frac}$ calculated over the ensemble of realisations developed with 5 random number seeds for each choice of the external pressure, $P_{ext}$. The lightly shaded region about each characteristic joining these data points represents the variation of $M_{frac}$ about its mean value. As can be seen from this plot, $M_{frac}$ increases steadily irrespective of whether the filament is of **Type I** or **Type II**. In general, however, the rate of cycling gas in to the dense phase is slower in **Type I** filaments than in the **Type II** filaments. Finally, while the fractional mass, $M_{frac}$, builds up gradually even for the initially super-critical **Type III** filament in a Solar-type environment (Case 5), it increases rapidly in a high-pressure environment (e.g., Cases 10 & 13). **Type III** filaments in general evolve relatively quickly. Evidently, higher external pressure facilitates cycling gas to higher volume densities. We will next explore if such high density filaments are also conducive to star formation.

Figure 5 (left) shows the temporal variation of the dense gas fraction based on column density for all simulations. Here, the dense gas fraction steadily increases to between 10% - 20% in filaments experiencing external pressure similar to that in the Solar -Neighbourhood (i.e., typically in the range of a few times $10^4$ $K$ $cm^{-3}$ - a few times $10^5$ $K$ $cm^{-3}$). For higher external pressures, i.e., upward of $\sim 10^6$ $K$ $cm^{-3}$, there is a rapid decline in dense gas fraction over time. This behaviour occurs irrespective of the linemass, because such filaments buckle (i.e., experience rapid distortion of their cylindrical geometry) as perturbations on their surfaces amplify rapidly. Indeed, this rapid decline in dense gas fraction could also help reconcile the inefficient nature of star

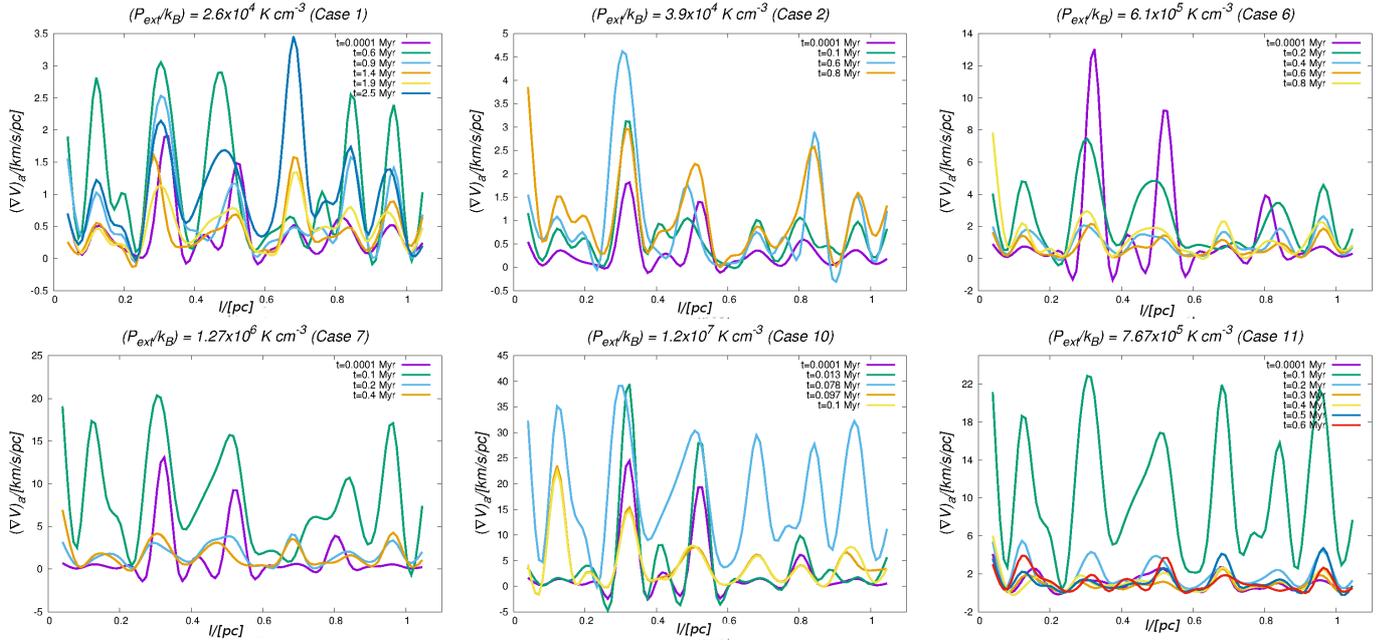

Figure 6.: Axial component of velocity gradient for filaments at different epochs of their evolution for different choices of external pressure. Given the identical nature of these plots and in the interests of brevity, we show here only a few cases such that they span the entire range of $P_{ext}$ & linemasses explored in this work. As with the plot on the lower panel of Fig. 2, these plots are also made from one of the simulations developed with 5 different random number seeds for each $P_{ext}$.

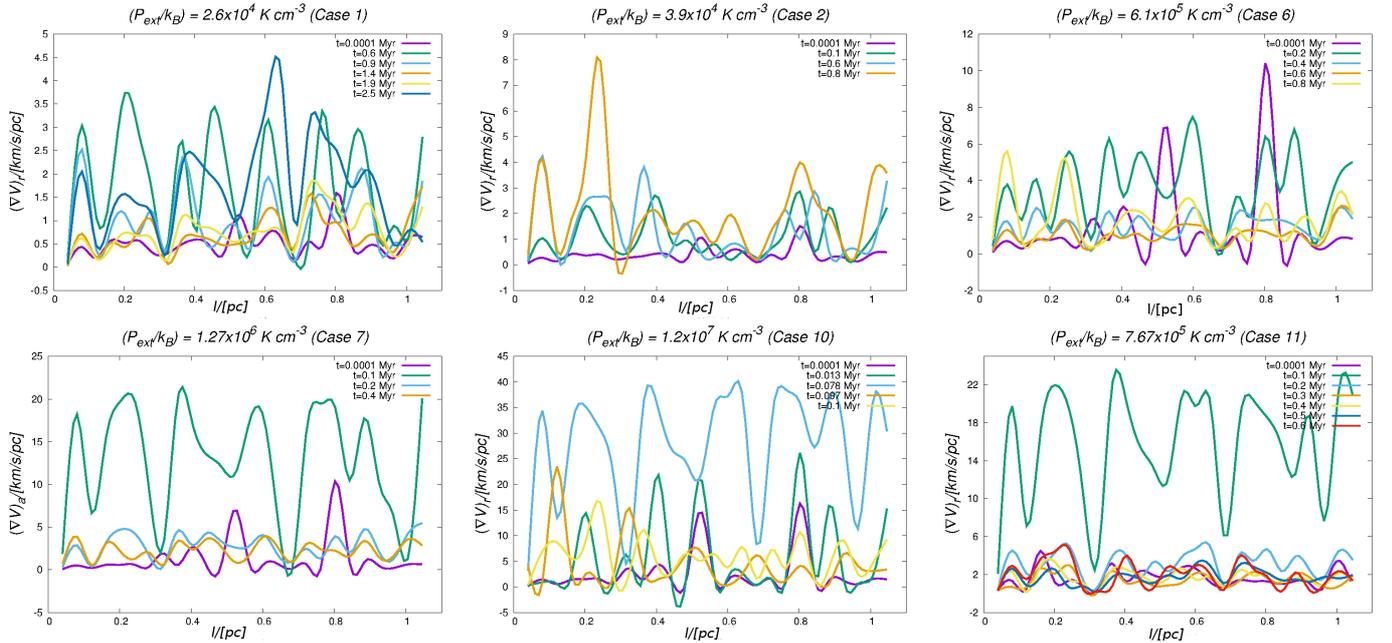

Figure 7.: Same as in Fig. 6 but now showing the radial component of velocity gradient.

formation in high-pressure environments.

This observed buckling of filaments in high-pressure environments is qualitatively similar to the evolution of shock-confined slabs assembled by gas-flows converging at super-sonic velocities. Such slabs are susceptible to the Shell Instability that manifests itself through rapid amplification of density perturbations on the slab surface due to the transfer of momentum between density crests and troughs, causing the shocked slab to buckle. This buckling motion of the slab (i.e., the filament in present simulations) cycles dense gas within it into the rarefied phase (e.g., Anathpindika et al. 2017).

In some of our earlier works we showed that while a higher external pressure may effectively cycle gas to higher densities, the fraction of putative star-forming gas actually decreases with increasing external pressure (Anathpindika et al. 2017, 2018). Now since a higher inflow velocity, $V_{inf}$, essentially means a larger external pressure, this implies that filaments with a higher rate of gas inflow buckle analogously to the behaviour of shocked slabs. So, while accretion enhances the filament mass, even trans-sonic accretion leading to pressures upward of $\sim 10^6 \ K \ cm^{-3}$ is likely to destabilise the filament and render inefficient the transfer of gas into the dense phase.

Finally, the filament in Case 5 was initially supercritical and yet the dense gas fraction there declines sharply on a relatively short timescale with external pressure comparable to that in the Solar Neighbourhood, i.e., $P_{ext}/k_B \sim 10^5 \ K \ cm^{-3}$. In this case, however, the filament contracts on a relatively short timescale and the formation of cores within it appears to be roughly simultaneous. As can be seen in the top left-hand panel of Fig. A3 the filament in this case contracts to form a thin dense spine which explains the low dense gas fraction. Interestingly, this case is the only one with relatively low external pressure in which the dense gas fraction decreases rapidly despite the steady contraction of the filament. In other cases (viz., Cases 6-10) where a similar decline in the dense gas fraction is also observed, the filaments buckled, segregating pockets of dense gas within them which causes the dense gas fraction based on column density to fall. These observed general trends in the temporal variation of the dense gas fraction based on volume density, and on the column density hold true even if $\mathcal{M}^{int} \gg 1$ for the respective filaments (as in cases 11, 12 & 13.) We revisit this point in §4.

Finally, Fig. 5 (right) shows the column density probability distribution function, i.e., the NPDF of the ten cases at the terminal epochs of their respective simulations. NPDF, the probability ($p_N$) that the column density lies between N and N + dN, is a useful diagnostic to determine the star-forming ability of MCs (e.g., Federrath & Klessen 2012, Kainulainen *et al.* 2009, Schneider *et al.* 2013, Stutz & Kainulainen 2015). NPDFs for MCs exhibit a density peak before turning over into the higher and lower density ends of the gas distribution. For typical star-forming regions, in fact, the NPDFs exhibit a well-defined power-law extension with a negative slope towards higher column densities (e.g., Goodman *et al.* 2009, Kainulainen *et al.*, 2013). That only a few filaments discussed here had begun forming stars (represented by sink particles) at the time respective simulations were terminated, is evident from the shapes of these NPDFs. Since we do not follow the evolution of individual cores that form in our filaments, it is unsurprising that these NPDFs do not exhibit power law extensions into higher column densities. Nevertheless, the NPDFs derived here are qualitatively similar to those reported for typical filamentary *Herschel* clouds.

### 3.2.2 QUASI-PERIODIC OSCILLATIONS OF THE VELOCITY GRADIENT

Periodicity in density & velocity structure along filaments has been known for quite some time now (e.g., Schneider & Elmegreen 1979; Dutrey *et al.* 1991). These suggestions have been confirmed further by several recent observations (e.g., Henshaw *et al.* 2020; Chen *et al.* 2020). While the determination of the amplitude of velocity fluctuations may be subject to instrumental sensitivity and uncertainties in estimating chemical abundances, the appearance of fluctuations themselves seems to be ubiquitous. So our aim in this work is merely to explore the impact of $P_{ext}$ on the periodicity & on the relative change in the amplitude of such fluctuations in the velocity field.

Figure 6 shows the axial component of the velocity gradient, i.e., $(\nabla V)_a$, along the lengths of the filaments in our simulations at different epochs of their evolution. We calculate the velocity gradients at different locations along the length of the filament first by determining the velocity peaks at these locations, i.e., in the plane of the filament axis, followed by calculating the difference in these velocity peaks over the distance between the locations. By *velocity peaks*, we mean here maxima in the velocities of gas particles in a slice taken at each location along the filament-axis and may be treated as proxies for the observationally determined line of sight velocities towards filaments. Gradients calculated in this manner are then resolved into their respective components in the axial direction along the filament-axis, i.e., $(\nabla V)_a$, and in the radial direction orthogonal to the axis, i.e., $(\nabla V)_r$.

Four crucial inferences can be drawn from these plots - **(i)** $(\nabla V)_a$ increases with increasing external pressure, or equivalently, with increasing velocity of the gas being accreted ($V_{inf}$); **(ii)** $(\nabla V)_a$ observed here is typically 2-3 times higher than those observed in Paper II for non-accreting filaments. Interestingly, the accreting gas does not damp out these quasi-oscillatory features of $(\nabla V)_a$. In fact, given that the trigger and subsequent amplification of these velocity fluctuations is qualitatively similar to the typical features of the shell instability (e.g., Vishniac 1983), it is likely that inflows with higher Mach numbers ($\mathcal{M}_{inf}$) will generate even larger velocity gradients, **(iii)** the ambient environment seems to have little impact on the periodicity of the fluctuations in the velocity-field, although not all of them grow, and **(iv)** the initially supersonic gas in the filament (i.e., Cases 11, 12 & 13) does not affect the peak amplitude of $(\nabla V)_a$ as it appears consistent with that observed in other simulations with comparable external pressure, but suffused with initially subsonic gas. Nevertheless, the initially supersonic gas seems to excite many more unstable modes as is evident from the higher periodicity of fluctuations in the velocity-field.

Plots in Fig. 7 are similar to those in Fig. 6 but now show the radial component of the velocity gradient, i.e., $(\nabla V)_r$. As with the axial component seen in the magnitude of $(\nabla V)_r$ also increases with increasing external pressure, i.e., with a higher inflow velocity ($V_{inf}$), though it is generally comparable to the axial component and like it, also exhibits oscillatory features. Recall that such oscillations in $(\nabla V)_r$ were also seen in the case of non-accreting filaments discussed in Paper II which suggests that oscillatory features in $(\nabla V)_r$ appear irrespective of whether a filament accretes gas. While we observe a higher magnitude of $(\nabla V)_r$ for a higher external pressure, $(\nabla V)_r$ peaks correlate more strongly with the external pressure than with the inflow velocity, be due to the limited sensitivity of their data, results presented here together with our earlier results in Paper II, however, suggest that the detection of oscillatory features in $(\nabla V)_r$ towards filaments is not a good proxy for detecting signatures of accretion. Indeed, such features can also appear in non-accreting filaments, a conclusion that is reinforced by the analytic work of Gehman *et al.* (1996), as well as by the observation here that $(\nabla V)_r$ peaks are more strongly correlated with $P_{ext}$.

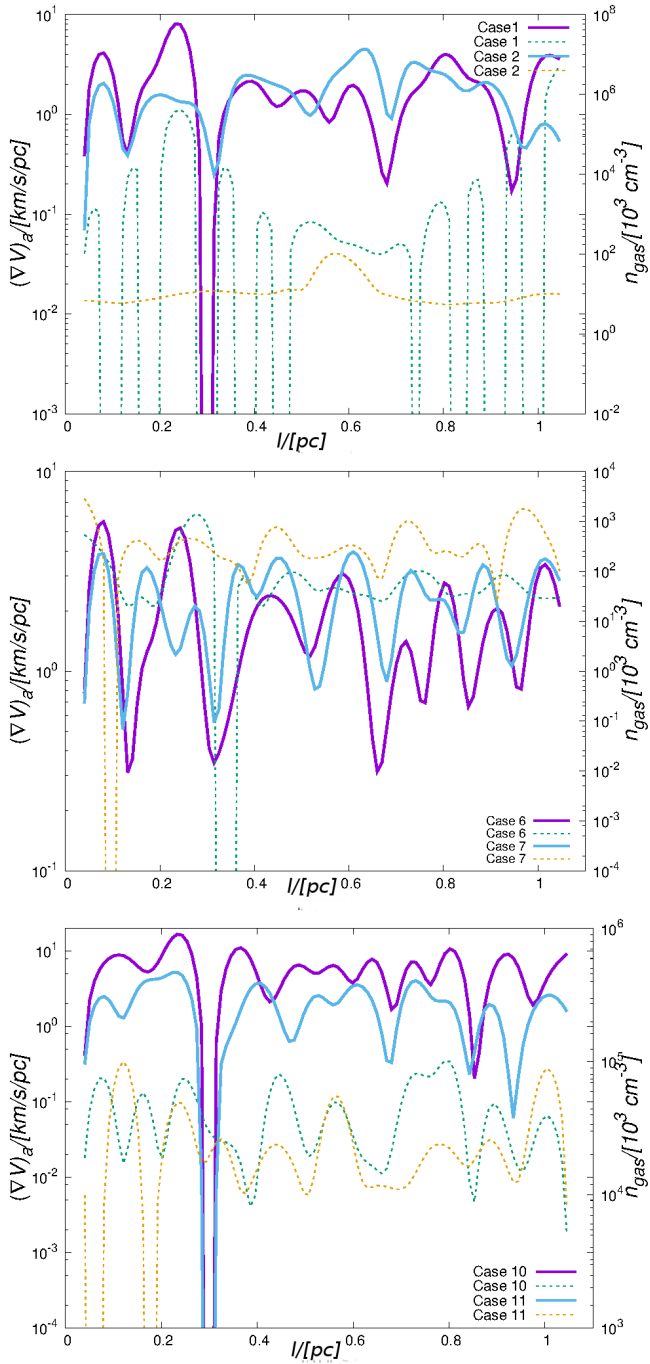

Figure 8.: Similar to plots in Figs. 6 & 7, but now shown for only the terminal epoch of the filaments in respective simulations and superposed with the location of density peaks along the filament-length.

Figure 8 shows the radial component $(\nabla V)_a$ with the density peaks overlaid on it. Continuous and dashed lines in these plots represent respectively the velocity gradient, and the density peaks in each plot. Evidently, the fluctuations of $(\nabla V)_a$ roughly correlate with the location of density enhancements along the length of the filaments. That these two quantities are indeed correlated is also reflected by the Pearson's Normalised cross-correlation which varies between ∼0.75 - 0.87 for these respective plots. In other words, the density and velocity peaks seem co-spatial, though not all perturbations condense out to form cores. The correlation between the density and velocity peaks is also visible in the filaments with the initially supersonic gas (i.e., Cases 11, 12 & 13). Perturbations in all our simulations are seeded purely by white-noise and the typical separation between them is,

$$\lambda_{sep} = \frac{22.1 a_{eff}}{(4\pi G \rho_c)^{\frac{1}{2}}} \quad (5)$$

(Nagasawa 1987); where $a_{eff}$ is the effective sound speed and $\rho_c$ is the mean central density of the filament.

For the typical values of physical parameters observed in these simulations, $\lambda_{sep} \lesssim 0.1$ $pc$, which in fact, is also comparable to the separation calculated according to the Gehman *et al.* analysis. The fact that only some perturbations condense out is especially true for the **Type I** filaments (i.e., plots corresponding to Cases 1, 3, 6, 8 & 11), and for the Type II filament in Case 12 with initially supersonic gas. Growth of other perturbations in these cases is stymied as they do not acquire sufficient mass which is the essence of the *collect-and-collapse* mode. *Evidently, occurrence of quasi-oscillatory features in the respective velocity gradients is merely indicative of the presence of spatial perturbations along the length of the filaments.*

### 3.2.3 EXTERNAL PRESSURE AND THE FILAMENT FULL WIDTH AT HALF MAXIMUM (FWHM$_{fil}$)

Figure 9 (left) shows the temporal variation of the FWHM$_{fil}$ of the filament for the set of simulations discussed above. Estimation of the FWHM$_{fil}$ has been described in detail in Paper I. Briefly, we calculate the FWHM$_{fil}$ by measuring the width of the lognormal distribution fitted to the column density across the cross-section of the filament at various locations along its length. If $\sigma_r$ is the dispersion of the corresponding estimates of the filament width, i.e., the average dispersion of these estimated filament widths, then FWHM$_{fil}$ $= 2(\sqrt{2\ln 2})\sigma_r$. While the FWHM$_{fil}$s for the **Type I** filaments in Cases 1 and 3 where the external pressure as reflected by Spearman rank coefficients ($S_n$) of 0.69 and 0.48, respectively. This behaviour suggests that $(\nabla V)_r$ is more sensitive to the external pressure than it is to the inflow velocity. Inclusion of the filaments with initially supersonic gas in Cases 11-13 does not significantly alter the $(\nabla V)_r - P_{ext}$ correlation.

Recently, Chen *et al.* (2020) also reported similar oscillatory features in the radial direction along filaments in the NGC 1333 region. They could not, however, conclusively identify any signatures of accretion onto the observed filaments. While this lack of evidence could is also low are relatively large, filaments suffused with initially supersonic gas in Cases 11-13 are the broadest in this ensemble of simulations and can be seen as the three outliers in this figure. It will be recalled from the plots in Fig. 2 that these are also the filaments with the biggest central radius, $r_{flat}$, and with a relatively large concentration parameter, $C$. In the remaining cases, however, we observe that the FWHM$_{fil}$ almost always progressively diminishes over time.

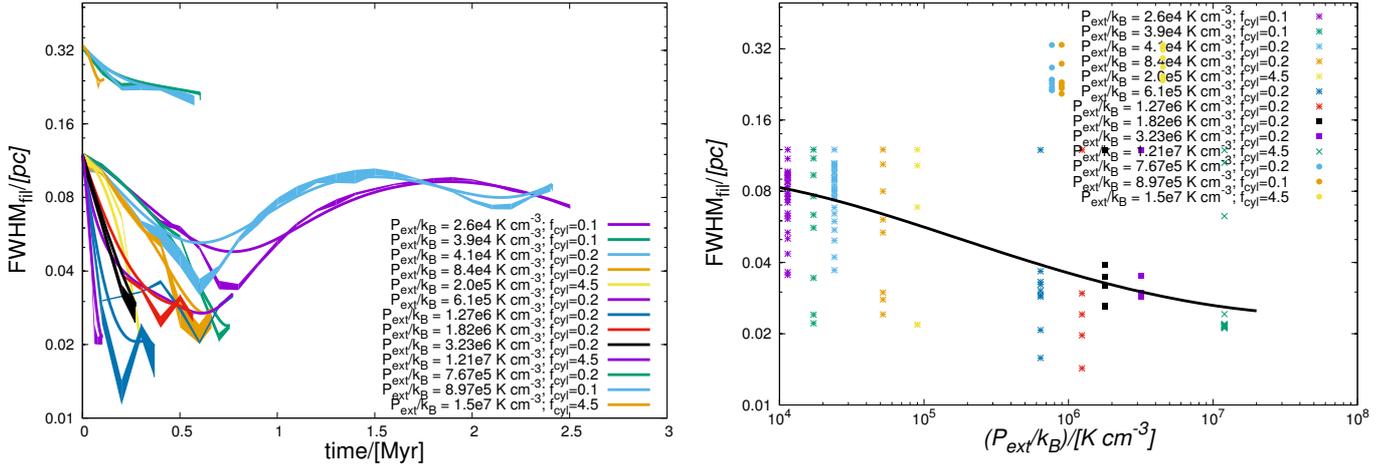

Figure 9.: *Left - panel:* Temporal variation of the FWHM$_{fil}$. As in the Figs. 2, 4 & 5 the lightly shaded region about each characteristic in either plot represents the variation of the FWHM$_{fil}$ over the simulations developed with 5 random number seeds for different choices of the external pressure. *Right - panel:* FWHM$_{fil}$ as a function of external pressure. Multiple data points for a given external pressure were generated by calculating the FWHM$_{fil}$ at different epochs of filament evolution. The thick black line on this plot represents the locus of points corresponding to the mean FWHM$_{fil}$ for each simulation. As before, individual data points here represent the mean value of FWHM$_{fil}$ across the simulations developed with 5 random number seeds for different $P_{ext}$.

Figure 9 (right) further illustrates that filament width progressively diminishes with increasing external pressure with the exception of the filaments in Cases 11-13. This correlation for the first 10 Cases is consistent with the analytic prediction by Fischera & Martin (2012). Evidently, the filament width is unlikely to be uniform across different environments. Thus it appears that only the filaments with subsonic gas in Solar -Neighbourhood exhibit widths on the order of ∼0.1 pc. Similarly, filaments with with subsonic gas in more hostile environments are likely to thinner, but also severely eviscerated. Conversely, those with supersonic gas are likely to be puffed up.

### 3.3 OTHER CORRELATIONS

The facts that accreting filaments contract radially to acquire a centrally peaked density profile and the accreted gas replenishes turbulence within them motivate these correlations. Figure 10 shows correlations between the column density of filaments, their linemass and the velocity dispersion (Virial parameter). As with similar plots discussed above, data points marked in various panels represent the magnitude of the corresponding physical quantity calculated at different stages of filament evolution. The top panel reveals that the filament linemass is rather poorly correlated with its column density (Spearman rank coefficient, $S_n$=0.08) even when the entire range of external pressure is considered. At relatively low external pressures (i.e., typically a few times $10^4$ $K$ $cm^{-3}$; e.g., Cases 1-5), however, this correlation is even worse, and indeed barely visible. The central panel of Fig. 10 shows the correlation between the velocity dispersion ($\sigma_{gas}$) and mean linemass from our simulations (Spearman rank coefficient, $S_n$=0.11). As with the previous correlation, however, this correlation also appears to be rather weak for lower external pressures of typically only a few times $10^4$ $K$ $cm^{-3}$.

Finally, the lower panel of Fig. 10 shows that the Virial parameter ($\alpha_{vir}$) weakly anticorrelates (Spearman rank coefficient, $S_n$= -0.08) with the fractional linemass ($f_{cyl}$) calculated at the mean temperature of the filament at that epoch. The Virial parameter, $\alpha_{vir} = \frac{2\sigma_{tot}^2}{GM_l}$, where $\sigma_{tot}^2 = \sigma_{gas}^2 + a_{gas}^2$, and $a_{gas}$ is the sound speed for the filament. Admittedly, we do not observe any meaningful correlation between these respective physical quantities. The vertical dashed line on this plot separates the thermally sub-critical filaments from those that are super-critical. Observe that most of our filaments irrespective of whether they are sub-critical/super-critical are virially bound ($\alpha_{vir}$ < 2), though a few of each are virially unbound. This observation is true even for the filaments initially suffused with supersonic turbulence, i.e., those in Cases 11-13. Admittedly, the boundedness of a filament must have little bearing on its ability to fragment and form putative star-forming cores. We observe that filaments in low pressure environments are typically thermally sub-critical, but virially bound. By contrast, filaments in Solar-type and in extreme environments such as in Cases 10 & 13 generally become super-critical, but can sometimes become virially unbound.

Respective plots in Fig. 11 illustrate that the col-umn density (left panel), and the velocity dispersion (right panel) both increase with external pressure, a result consistent with the analytic prediction by Fischera & Martin (2012). The velocity dispersion in the current simulations is the result of a combination of momentum injected by the accreting mass as well as the radial contraction. In fact, as with the respective components of $(\nabla V)$, the velocity dispersion in the present simulations is also typically a factor of 2-3 higher than that reported in Papers I & II. Accretion of mass is also why we observe a much clearer trend of increasing column density with increasing external pressure. In Paper I where filaments remained relatively starved of mass, however, the dependence of column density on external pressure was weak even in high-pressure environments.

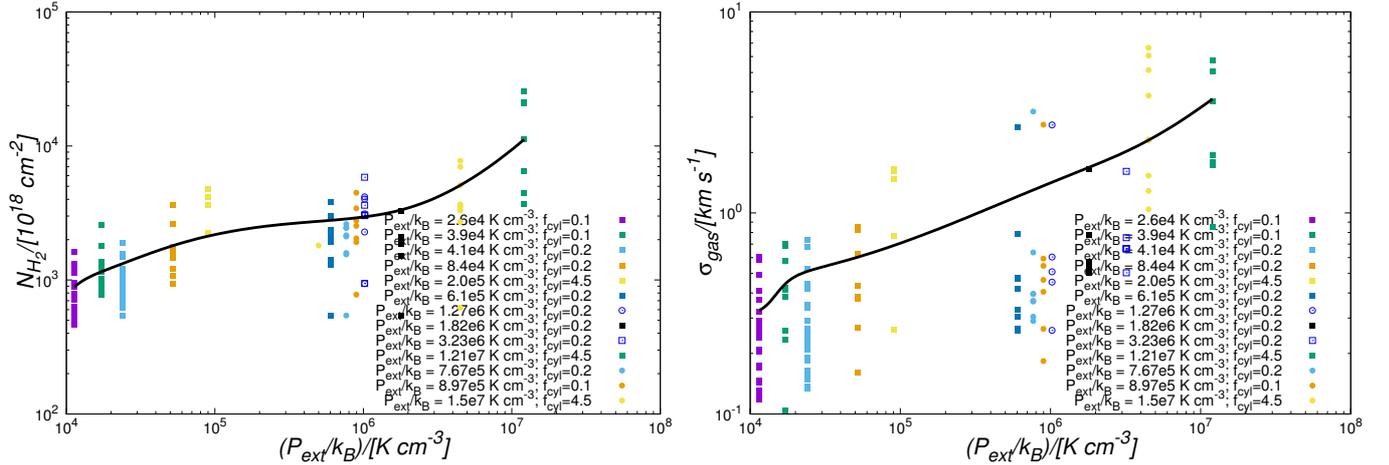

Figure 11.: Similar to the plots in Figs. 9 & 10, but now showing respectively the column density, and the velocity dispersion as a function of the external pressure.

## 4 DISCUSSION

We observe that accreting filaments evolve through a series of radial contractions and expansions, i.e., a series of inward and outward gas-motion in a circular direction perpendicular to the filament-axis, irrespective of their linemass and external pressure. Filaments in this work become radially unstable due to a combination of dynamical cooling, turbulent dissipation and accretion of gas that increases filament linemass. Irrespective of the external pressure, the filaments develop truncated density profiles. Simulations in this work show that the interplay between turbulence, thermal pressure & self-gravity can generate shallow density profiles. This behaviour is also consistent with the findings of the analytic work by Gehman *et al.* (1996) as well as some of our earlier numerical work (Anathpindika & Freundlich 2015).

These results show that filaments in low-pressure environments (i.e., $\frac{P_{ext}}{k_B} \lesssim 10^4\ Kcm^{-3}$) and those with initially supersonic gas must have a bigger inner radius (and by extension, a higher concentration parameter $C$), and develop a shallow density profile with the Plummer exponent $p \sim 2$. With the exception of the filaments in Cases 11 - 13 that were superposed with initially supersonic gas, filaments become thinner and have smaller inner radius (and by extension, a lower concentration parameter $C$), with increasing pressure. This latter behaviour, which is also consistent with the results from the semi-analytic work by Fiege & Pudritz (2002), can be seen from the plots shown on various panels of Fig. 2. Indeed, the broad filament in Case 13 is consistent with the findings of Federrath *et al.* (2016) about filaments in the *Brick*.

We note that filament-evolution in all our simulations, including that in Cases 11 - 13, is gravity dominated. Filaments in the latter, i.e., in Cases 11 - 13, are especially interesting because their evolutionary cycle is remarkably different from the turbulence dominated *fray and fragment* mode observed by Clarke *et al.* (2017) in those that are initially suffused with supersonic turbulence. In purely isothermal simulations, Clarke *et al.* observed the formation of fibrous sub-structure within filaments irrespective of whether they were dominated by the compressional or the solenoidal mode of turbulence. We note, however, that unlike the pressure-confined filaments in this work, the accreting filaments modelled by Clarke *et al.* were essentially placed in vacuum. Furthermore, $\dot{M}_{acc}$ assumed in the simulations discussed here is a factor of 6 to 11 smaller than that assumed by Clarke *et al.* meaning that the turbulence within our filaments is not sufficiently replenished over the course of their evolution.

Various rendered images in Figs. A1 - A3 show that accreting filaments in low-pressure environments evolve on relatively long timescales which are comparable to, or greater than the so-called *e-folding* ($e_{fold}$) timescale. The so-called geometrical mode of filament fragmentation identified by Gritschneder *et al.* (2017) and Heigl *et al.* (2018), which manifests itself through the growth of density perturbations, can be seen in the **Type I** (i.e., the initially sub-critical filaments that remain so even after accreting gas, $f_{cyl} < 1$) filaments of Cases 1, 3, 6, 8 & 11. This kind of behaviour is also observed in the **Type II** filament in Case 12 that was initially suffused with supersonic gas.

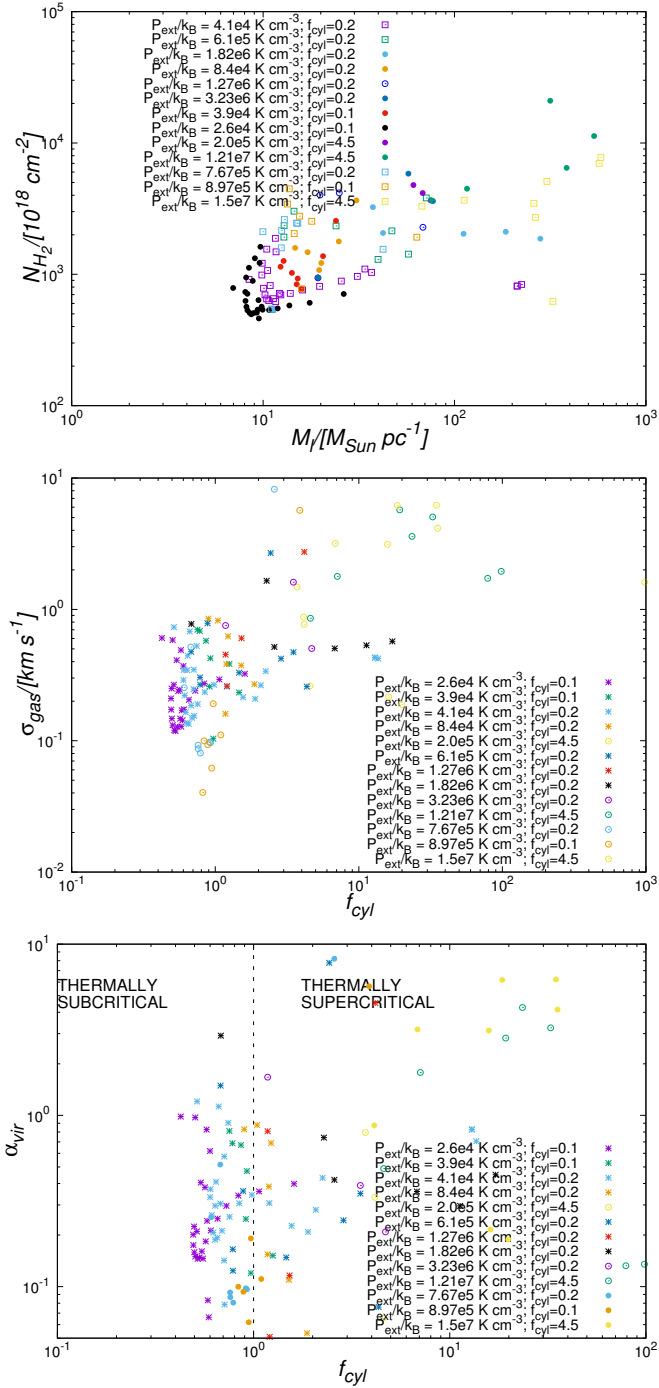

Figure 10.: Similar to the plots in Fig. 9, but now showing respectively the column density, velocity dispersion, and the Virial parameter are shown on the top, central & bottom panels.

The filament in these respective simulations becomes unstable to the *sausage-type* instability (also sometimes referred to as the deformation instability; Nagasawa 1987), so that broad condensations (i.e., mean aspect ratios in the range 1.05 - 1.13 for those in Cases 1, 3, 6, 8, 11 & 12 respectively) begin to appear along the length of the filaments in respective cases. Some of these condensations merge, while a few survive and become massive enough to collapse (i.e., the *collect-and-collapse* mode). We also observe that the evolutionary timescale of the filament becomes significantly smaller with increasing external pressure. Thus the filaments in Cases 6, 8 & 12, where the external pressure is in the range of a few times $10^5$ $K$ $cm^{-3}$ and a few times $10^6$ $K$ $cm^{-3}$, respectively, evolve on less than half the timescale of the filaments in Cases 1 & 3 where $P_{ext}/k_B \sim 10^4$ $K$ $cm^{-3}$.

Gas accreted by filaments in Cases 6 & 8 is warm, a situation typically envisaged towards the Galactic CMZ). The **Type I** filaments in these respective cases also fragment via the *collect-and-collapse* mode, but much faster than in Cases 1 & 2 where $P_{ext}$ is at least 2 orders of magnitude lower. Also, the filament becomes severely ablated in such hostile environs (as in Case 8, for instance). In these two cases the thermal pressure due to the accreted warm gas initially provides buoyancy until it gradually cools and so, the contractional wave driven by the incoming gas is not strong enough to enhance the central density sufficiently so that compressional *Jeans-type* fragmentation can ensue. In other words, the *collect-and-collapse* mode of fragmentation must generally dominate weakly self-gravitating filaments.

This observation could help us reconcile the well-separated *broad* cores in the *Musca* filament (e.g., Hacar *et al.* 2016) that look like pearls on a string. *Collectively, these findings suggest that sub-critical filaments are more likely to form cores via the collect-and-collapse mode in low to intermediate pressure environments like that in the Solar-Neighbourhood. In any case, star-formation in such filaments is likely inefficient, irrespective of the external pressure.* It is therefore clear that irrespective of the linemass, the external pressure does in fact bear upon the evolution of accreting filaments, their density distributions, their inner filament radii and the morphologies of cores that they spawn. It is also important to note that the filaments in Cases 11 - 13 initially suffused with supersonic turbulence also evolve in a manner similar to those in the first 10 Cases where they were suffused with initially subsonic turbulence. The only difference is that the initially turbulent filaments are fluffier than the latter. It remains to be seen how the ambient environment affects the filament accretion, and subsequently, its linemass. We propose to investigate this problem in a more dynamic set-up where filaments are allowed to assemble self-consistently.

### 4.1 STAR FORMATION EFFICIENCY

There is no dearth of literature showing considerable variation in star formation efficiency across molecular clouds (e.g., Lee *et al.* 2016; Vutisalchavakul *2016*). The suggestion that external pressure modulates the efficiency (i.e., the fraction of gas in a volume converted to stars) with which gas is converted into stars has been gaining traction in recent years (e.g., Hughes *et al.* 2010, 2013; Meidt et al. 2013). Recent numerical work on this subject shows that clouds in high-pressure environments (i.e., $P_{ext}/k_B \gtrsim 10^6$ $K$ $cm^{-3}$) are inefficient at cycling gas into the dense, putative star-forming phase (e.g., Anathpindika *et al.* (2017; 2018). Equivalently, a relatively poor efficiency of cycling gas into the dense phase could reconcile the observed inefficiency of star-formation in such environments.

Plots in Fig. 4 show that the dense gas fraction by volume ($M_{frac}$) rises during the contractional phase of filaments before decreasing rapidly during the later phases of evolution when they expand, and that the fraction of gas cycled to higher volume densities increases with increasing external pressure. Crucially, however, the cycling of an increasing fraction of gas to higher volume densities does not necessarily mean a higher dense gas fraction by column density, as is evident from the plot on the left-hand panel of Fig. 5. This apparently contradictory situation arises because the rapid amplification of perturbations on the surfaces of filaments, especially in high-pressure environments (i.e., pressures upward of $\gtrsim 10^6$ $K$ $cm^{-3}$) causes them to buckle, and sometimes even rupture (i.e., actually broken in to several pieces), which quickly diminishes the dense gas fraction. In the process, filaments may even become severely eviscerated. For example, the filament in Case 9 loses about 40% of its mass by the time the simulation was terminated. The lost mass of course remains in the computational domain and becomes part of the diffuse medium surrounding the filament. The dense gas fraction by column density in Cases 11 & 12 barely reaches a maximum of ~25% before eventually falling off to ~10% at the time of termination of the respective simulations; Case 13 is even worse in regard as the dense gas fraction just about reaches ~12% before tapering off by the time calculations were terminated.

The column density PDFs (i.e., NPDFs) for filaments shown in Fig. 5 (right) are broadly consistent with those found for filamentary *Herschel* clouds (e.g., Kainulainen *et al.* 2009; Stutz & Kainulainen 2015). They are also consistent with the semi-analytic calculations of Myers (2017) who deduced similar NPDFs for elongated clouds modelled either as a 2-D truncated Plummer cylinder, a truncated prolate spheroid, or a stretched truncated prolate spheroid. It can be seen from this plot that the NPDFs of filaments experiencing external pressure typically upward of ~$10^6$ $K$ $cm^{-3}$ decline rapidly at higher column densities. Similarly, the NPDFs of filaments experiencing lower pressures (typically $\lesssim 10^4$ $K$ $cm^{-3}$) exhibit long power-law tails at lower column densities ($\lesssim 10^{21}$ $cm^{-2}$). This latter observation suggests that a considerable fraction of gas in such filaments must remain diffuse. Hence, they are likely to be inefficient at forming stars, a result consistent with the observational findings of e.g., Goodman *et al.* (2009), in respect of the diffuse clouds (i.e., column densities $\lesssim 10^{21}$ $cm^{-2}$).

Furthermore, the NPDFs for filaments experiencing intermediate range of pressures (between typically $\gtrsim 10^{5-6}$ $K$ $cm^{-3}$) peak at column densities $\gtrsim 10^{21}$ $cm^{-2}$, but exhibit only relatively short power-law tails at low column densities. Such filaments are therefore likely most efficient at forming stars. Interestingly, however, the NPDFs for the **Type I & II** filaments in Cases 11 & 12 respectively, though qualitatively similar to the others at comparable external pressure, are clipped at the high density end. Also, the peaks of these respective NPDFs are shifted towards lower densities. This observation is not altogether surprising since turbulence inhibits the process of cycling gas into the dense phase.

Having lost a substantial fraction of its mass over the course of its evolution, the truncated NPDFs at higher densities for the **Type III** filaments in **Cases 10 & 13** again highlight the fact that only a relatively small fraction of its mass is ever cycled to column densities $\gtrsim 10^{21}$ $cm^{-2}$. This observation reinforces the fact that star-formation must be inefficient in high-pressure environments. While it is clear that the ambient environment modulates the star-forming ability of clouds, it is also interesting to note that Kainulainen *et al.* (2013) argued that variations in the strength of the compressional component of turbulence affect the fraction of the putative star-forming gas, as reflected by the slopes of NPDFs towards the high density regime.

### 4.2 QUASI-PERIODIC OSCILLATIONS

Schneider & Elmegreen (1979) and Dutrey *et al.* (1991) were among the first to report the existence of periodic structure along filaments. Density fluctuations in the ISM are seeded by the interplay between various dynamic instabilities and putative star-forming regions are assembled in regions of converging gas-flows (see review by, e.g., Dobbs & Baba 2014). In an earlier semi-analytic work, Gehman *et al.* (1996) showed that fluctuations in the density and velocity fields could be generated due to the propagation of waves and instabilities and that there is an approximate correspondence between the two fields, unlike other works (e.g., Nakamura *et al.* 1991), which suggested magnetohydrodynamic motions like Alfvén waves as the possible sources of these observed fluctuations in the density/velocity-fields.

Gehman *et al.* particularly showed that pressure-confined filaments are susceptible to a class of wave-solutions that propagate along their axes, and in the direction orthogonal to their axes. They further showed that these waves have a characteristic fragmentation length-scale, which in the case of purely logatropic filaments depends on external pressure and varied as $P_{ext}^{1/2}$. Their models suggest that the density and the velocity fields are approximately cospatial. They also predict a converging velocity-field towards density peaks, but a non-convergent velocity-field in regions of stable modes (i.e., those that do not grow).

Recent observations of the ISM by Henshaw *et al.* (2020), for example, show that fluctuations in the velocity-field are ubiquitous. Analysis of observational data spanning a wide variety of Galactic environments by these authors further showed that the periodicity of velocity structure was often similar to that of the underlying density structure. They therefore conjectured that fluctuations in the density and velocity-fields were likely the key to understanding the origin of gas-flows that assemble star-forming regions in the ISM. On smaller scales, Chen *et al.* (2020) similarly observed remarkably coherent velocity fluctuations in the NGC 1333 region where they found that the magnitude of the radial component of the velocity gradient decreased towards the spines of observed filaments and inferred them as signatures of ongoing accretion by filaments. Chen *et al.*, however, did not detect any strong evidence to suggest large scale accretion by the observed filaments.

Similar velocity gradients, typically on the order of a few $km$ $s^{-1}$ $pc^{-1}$, were also reported in the Serpens South star-forming region (e.g., Kirk *et al.* 2013; Fernández-Lopéz *et al.* 2014), and the Serpens Main star-forming region (e.g., Lee *et al.* 2014). Fluctuations in the velocity field were also reported in the *Brick* and the Sgr B2 region in the Galactic CMZ. Given the extreme ambient environment in the CMZ, peaks of the fluctuations in the centroid velocities are typically an order of magnitude higher than that for clouds in the local neighbourhood (e.g., Henshaw *et al.* 2016; 2019). In fact, Henshaw *et al.* (2016) also observed that the fluctuations in the velocity field were approximately co-spatial with the massive ($\sim 10^4$ $M_\odot$) clouds in the region which led them to suggest that the observed velocity fluctuations were likely triggered by the gravitational instability.

Still more recently, Wallace *et al.* (2022) observed velocity gradients on the order of a few $km$ s$^{-1}$ towards the Sgr E region which itself is believed to be at the turbulent intersection of the dust lane associated with the Galactic bar and the CMZ and may also be entering the CMZ. Evidently, the ambient environment must affect the peaks of observed velocity fluctuations. Observations of filaments at different length scales show some evidence of larger axial components of the velocity gradient on small spatial scales, with perhaps $(\nabla V)_a$ scaling as the inverse of filament length. See for e.g., the review by Hacar *et al.* (2022).

Various panels of Figs. 6 and 7 show that quasi-periodic oscillations in the velocity gradient along the axis of the filament, and in the direction orthogonal to it are ubiquitous in the sense that such features are visible irrespective of the magnitude of external pressure and filament linemass. Our findings here are therefore consistent with similar observational inferences suggesting that velocity and density fluctuations are likely ubiquitous as they are visible across a wide range of spatial scales, even up to a few parsecs (e.g., Henshaw *et al.* 2020). Evidently, the peaks of these oscillations increase with increasing external pressure which is consistent with observational inferences. Various panels of Fig. 8 also suggest that these oscillations overlap with the spatial separation of fragments that form in the respective filaments. This observation is reinforced by the fact that Pearson's Normalised correlation coefficients for these plots is closer to unity. Noticeably, however, not all the fragments condense to form cores that collapse. Taken together, Figs. 6 - 8 along with our observations of similar oscillatory features in the case of non-accreting filaments examined in Paper II, make it clear that such fluctuations in the velocity-field can be triggered even by purely hydrodynamic instabilities.

### 4.3 FILAMENT WIDTH

The idea that filaments, at least those in the nearby molecular clouds, have a universal width, i.e., FWHM$_{fil}$ $\sim 0.1$ $pc$ (e.g., Arzoumanian *et al.* 2011, 2013; Andre *et al.* 2016), has recently come under increasing scrutiny (e.g., Panopoulou *et al.* 2014, 2022). Convergence tests by André *et al.*(2022) for filaments in the Solar neighbourhood, however, show that the conclusion about the FWHM$_{fil}$ being $\sim 0.1$ $pc$ is robust. The authors therefore call for a theoretical framework to reconcile this observed lengthscale. That filaments in this work exhibit FWHM$_{fil}$ variations over the course of their temporal evolution is readily clear from the plots shown in Fig. 9 (left). Filaments suffused with initially supersonic gas, however, have the largest FWHM$_{fil}$ and are the three outliers in this figure. Furthermore, the decrease in filament width with increasing external pressure, with the exception of the filaments in Cases 11-13, is also evident from Fig. 9 (right). Indeed, thinner filaments with a FWHM$_{fil}$ significantly smaller than $\sim 0.1$ $pc$ have also been observed elsewhere in the Galaxy, as noted in §1 above. See also a more recent review by Hacar *et al.* (2022).

In their semi-analytic work, Gehman *et al.* (1996) showed that turbulence-supported logatropic filaments are generally wider and exhibit shallower density profiles in comparison with isothermal filaments. So the observations in Cases 11 - 13 here are consistent with the analytic predictions by Gehman *et al.*. More recently, Priestley & Whitworth (2021) showed numerically that filaments having widths on the order of $\sim 0.1$ $pc$ can be generated if the gas being accreted is initially mildly supersonic and has a Mach number, $\mathcal{M}_{inf} \lesssim 3$. They further argue that accretional flows with higher $\mathcal{M}_{inf}$ generate thinner filaments, and that the width of a filament is determined by the location of the accretion shock within a filament.

Indeed, Fig. 3 shows the filament width in these simulations is associated with the location of the inwardly propagating compressional disturbance. Filaments in low-pressure environments (i.e., similar to the Solar -Neighbourhood) typically have FWHM$_{fil}\sim 0.1$ $pc$. For higher external pressures, however, the filaments (with the exception of those in Cases 11-13) become thinner, a result consistent with the conclusion drawn by Priestley & Whitworth (2021). *Despite $\mathcal{M}_{inf}$ being a dimensionless empirical physical quantity, a given Mach number could correspond to a range of external pressures depending on the temperature of the inflowing gas. Even in this work, a transonic inflow (i.e., $\mathcal{M}_{inf} = 1$), generates $P_{ext}$ in the range $10^4 - 10^7$ $K$ $cm^{-3}$. So we believe that the external pressure is a better proxy for the ambient environment. Our conclusion here about $P_{ext}$ having a significant bearing upon the FWHM$_{fil}$ of filaments is therefore robust. Our results suggest that the filament width typically signifies the lengthscale on which it acquires a state of dynamic equilibrium.*

### 4.4 OTHER FILAMENT PROPERTIES

Surveys of filaments across a number of clouds in the local Solar-Neighbourhood have shown a direct correlation between the velocity dispersion of gas in filaments, $\sigma_{gas}$, their linemass, and the column density (e.g., Arzoumanian *et al.* 2013). While $\sigma_{gas}$ may indeed represent the strength of the underlying turbulent velocity field, it also includes contribution due to systemic gas-flows as in a contracting cylinder or in regions of localised collapse. In general, however, $\sigma_{gas}$ is a good proxy for turbulence, at least before individual density perturbations collapse.

Such correlations support the thesis that the gas accreted by filaments is a source of turbulence within them, while the accreted mass simultaneously increases their column density. Indeed, Fig. 10 (top & central panel) shows that such a correlation, albeit a very weak one, is visible in our own simulations, especially those where the respective filaments experience intermediate to high large external pressure, i.e., $P_{ext}/k_B \gtrsim 10^5$ K cm$^{-3}$. At lower pressures, however, there is no meaningful correlation. A similar trend between $\sigma_{gas}$ and the linemass is also seen in the plot in Fig. 10 (central panel).

Figure 10 (lower) shows that the Virial parameter weakly anticorrelates with the linemass irrespective of whether the filaments are sub-critical or super-critical. This anticorrelation is not as strong as that observed by Arzoumanian *et al.* (2013) for their sample of filaments. We note, however, that most of the filaments in our present work are virially bound. This is also true of the filaments in Cases 11 - 13 where they were initially suffused with supersonic turbulence, though over the course of their evolution they do sometimes become unbound. Generally speaking, the correlations observed in this work are rather weak which could be due to the idealised nature of our set-up.

Finally, Fig. 11 shows the correlations between the column density, $N_{H_2}$, and external pressure, and $\sigma_{gas}$ and external pressure. The former correlation is, however, somewhat weaker than that predicted analytically by Fischera & Martin (2012). For the latter, we suggest a higher accretional velocity injects stronger turbulence that manifests itself in the form of a higher velocity dispersion. This latter result corroborates recent observational findings that show a larger coefficient of the size-line width relation towards high-pressure environments (e.g., Rice *et al.* 2016). That the pressure due to turbulence, $P_{turb} (\equiv \rho_{gas}\sigma_{gas}^2)$, within the filament correlates tightly with the external pressure, $P_{ext}$, can indeed be seen from the plot in Fig. 12.

## 5 CONCLUSIONS

The principal conclusions of this paper are as follows -

1. The magnitude of external pressure definitely bears on the evolution of filaments and on the morphology of cores that spawn within them, irrespective of their linemass. Thus, sub-critical filaments in low to intermediate pressure environments evolve slowly to form broad cores via the *collect-and-collapse* mode.

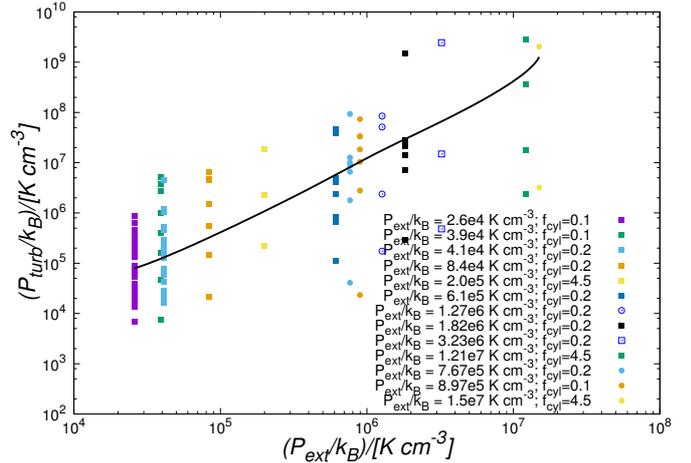

Figure 12.: Like the earlier plots in Figs. 9, 10 & 11 but now showing the turbulent pressure, $P_{turb}$, as a function of the external pressure. The black continuous line represents as before the locus of points corresponding to the mean $P_{turb}$ for each simulation.

2. With the exception of the filaments suffused with initially supersonic gas that are generally wider, there is a clear trend of decreasing filament width with increasing pressure and filaments in Solar Neighbourhood-like environments have widths typically on the order of $\sim 0.1$ *pc*.
3. Filaments at either extremes of pressure, i.e., towards the Galactic centre in the Central Molecular Zone where $P_{ext}/k_B \gtrsim 10^7$ $K$ $cm^{-3}$, or towards the fringes of the Galactic disk where $P_{ext}/k_B \lesssim 10^4$ $K$ $cm^{-3}$, are inefficient at cycling gas into higher densities and thus forming stars.
4. We observe quasi-periodic oscillatory features of the velocity gradient in all our simulations and peaks of their fluctuations generally increase with increasing pressure. Fluctuations in the density and the velocity-fields are also roughly cospatial. We suggest that oscillatory features of $(\nabla V)_r$ are a poor proxy for detecting signals of accretion on to filaments. In fact, we observe that $(\nabla V)_r$ correlates more strongly with the external pressure than with the inflow velocity of gas being accreted by the filament.
5. Filaments initially suffused with supersonic turbulence have physical properties generally similar to those suffused with subsonic turbulence, except that the former are much less efficient at cycling gas into the dense phase as reflected by a lower dense gas fraction, but have a bigger inner radius and by extension, a bigger FWHM$_{fil}$.
6. Filaments have higher column densities and higher velocity dispersion (also reflected by a higher pressure due to turbulence within the filament) with increasing external pressure, showing us that external pressure modulates properties of gas even on the scale of individual filaments. We also observe that the filaments modelled here are largely virially bound irrespective of their linemass. Filaments in a Solar-type and in extreme environment as in Cases 10 & 13 can, however, sometimes appear to be unbound.

## Acknowledgements

This project was initiated with funds made available under the From GMCs to stars project (GMCS/000304/2014), funded by the Department of Science & Technology, India. Simulations discussed in this work were developed using supercomputing facilities made available by Digital Research Alliance of Canada (https://alliancecan.ca) and Compute Canada Calcul Canada (www.computecanada.ca). The authors gratefully acknowledge useful discussions with Adam Ginsburg and Cara Battersby. The authors are grateful to anonymous referees for some critical comments that made the original manuscript much clearer.

## Data Availability Statement

No valid data repositories exist as the data generated by numerical Simulations discussed in this work are too big to be shared. Instead we discuss in detail the numerical methods and the initial conditions used to generate these data sets. The initial conditions file and the script of the numerical code can be made available to bonafide researchers on reasonable request.

## A Filament evolution

The **Type I** (i.e., $f_{cyl} < 1$, even after accreting gas) filament in a low pressure environment evolves over a longer timescale and forms cores through the *collect and collapse*[3] mode described in Papers I & II. This behaviour is also visible from the panels specifically corresponding to Cases 1 and 3 **(and also in Case 11)** in Figs. A1 and A2 **(and in Fig. A3, lower panel)**, respectively. Interestingly, there is also evidence of mergers between some fragments, as reported by Inutsuka & Miyama (1997). On the contrary, **Type II** (i.e., $f_{cyl} > 1$, after accreting gas) filaments, with the exception of that in Case 12, form cores via *Jeans - type* fragmentation. This mode of fragmentation can be readily seen in the panels specifically corresponding to Cases 2, 4 & 6 in Figs. A1-A3 respectively.

In contrast, the continually super-critical (**Type III**) filaments evolve rapidly to form spines, i.e., thin filamentary structures with centrally peaked density profiles, with the exception of that in Case 13. While fragmentation of the filament into distinct pinched cores is visible in Case 5 (e.g., in the picture on the top left hand panel of Fig. A3), no such fragmentation and subsequent formation of cores is visible in the filament in Case 10 (e.g., in the top right hand panel of Fig. A3), whereas the filament in Case 13 is significantly wider than that in Cases 5 & 10. At any rate the filaments in Cases 10 & 13 appear to be composed of a number of dense strands similar to fibres. There is, however, no evidence of core formation in these fibres or in the filament itself.

Rendered density images on lower panels of Fig. A3 similarly show the terminal epoch of the filament in respectively Cases 6, 11, 12 & 13. Observe that as with the filaments superposed with initially subsonic turbulence (i.e., in the first 10 cases), even those with supersonic turbulence (i.e., Cases 11, 12 & 13) evolve via radial contraction followed by axial fragmentation. In other words, the evolution of these latter filaments is still gravity-dominated. This is because as the initial turbulence dissipates, the critical linemass decreases and the filament is readily overwhelmed by an inwardly propagating compressional wave. It means, the inflowing transonic gas in this instance does not sufficiently replenish the turbulence. Thus the choice of the internal Mach number ($\mathcal{M}^{int}$) makes no qualitative difference to the evolutionary cycle of filaments which is not particularly surprising given its inefficient replenishment by the inflowing gas.

Respective images corresponding to Cases 6 & 11 on the lower panel of Fig. A3 make an interesting comparison as the filaments in either case are not only of the **Type I** kind, but the external pressure is also comparable. Evidently, the turbulence supported filament in Case 11 evolves on a timescale almost twice as long as that in Case 6. More importantly, the cores in the former simulation (i.e., Case 11) are broader than the natal filament and appear to have formed via the *Collect and Collapse mode*. On the contrary, the filament in Case 6 appears to have spawned cores via the *Jeans - type* fragmentation.

The **Type II** filament in Case 12 where the external pressure is comparable to that in Cases 6 & 11, also evolves like that in the Case 11 and appears to have formed cores via the *Collect and Collapse mode*. This is unlike the other Type II filaments in Cases 2, 4 & 7 where cores form via the *Jeans-type* mode of fragmentation. The filaments in Cases 2 & 4 are shown in Figs. A1 & A2. These observed differences in filament evolution suggest that weakly self-gravitating filaments must fragment via the *Collect and Collapse* mode. The continually super-critical **Type III** filament in Case 13 evolves similarly to the one in Case 10, though it is fluffier and is also more profusely striated than the one in Case 10.

---

[3] where a density perturbation accumulates gas and collapses only after becoming sufficiently massive

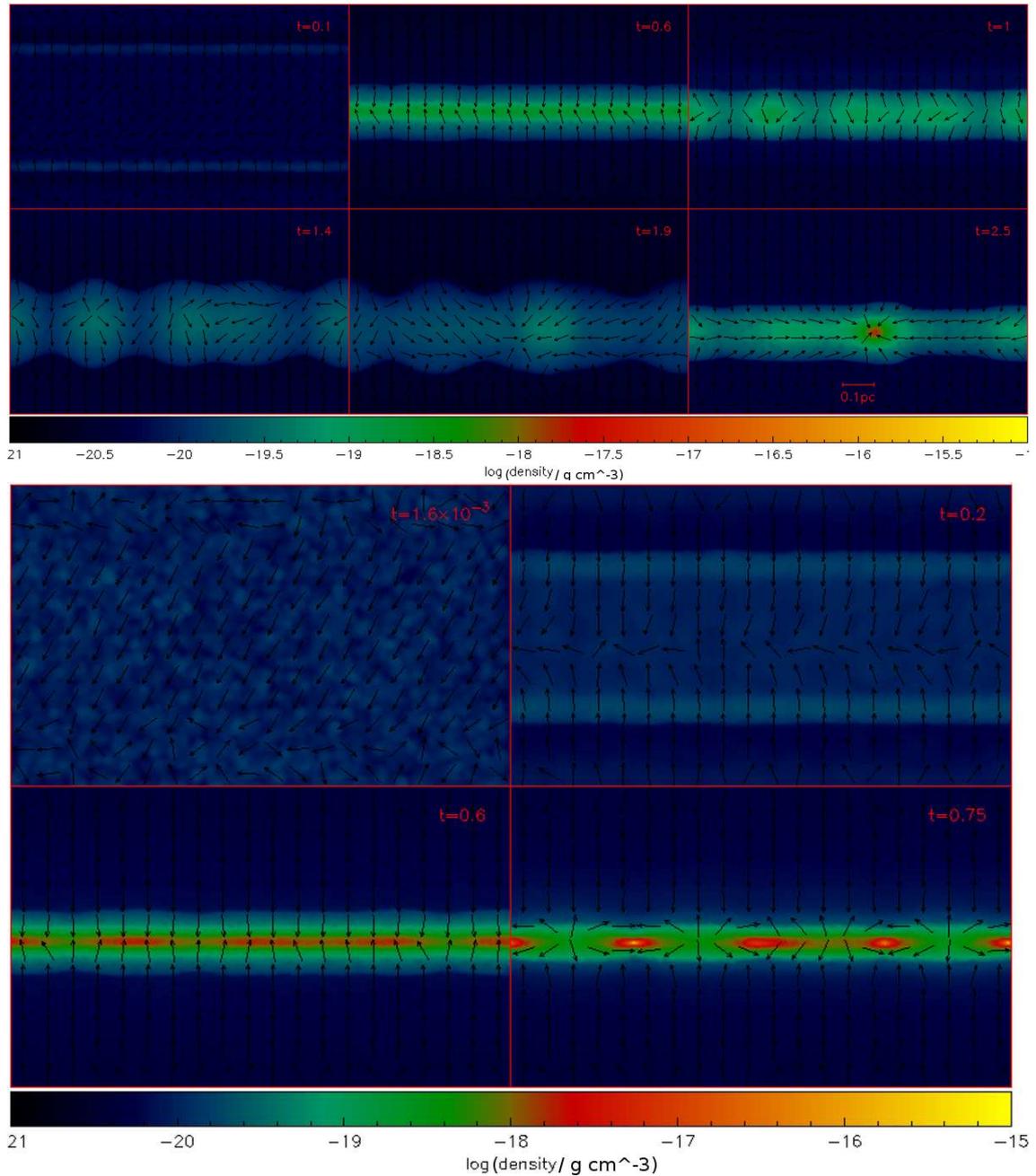

Figure A1.: Rendered density images showing the evolutionary sequence of the filaments in respectively Cases 1 (*Upper panel*), and 3 (*Lower panel*).

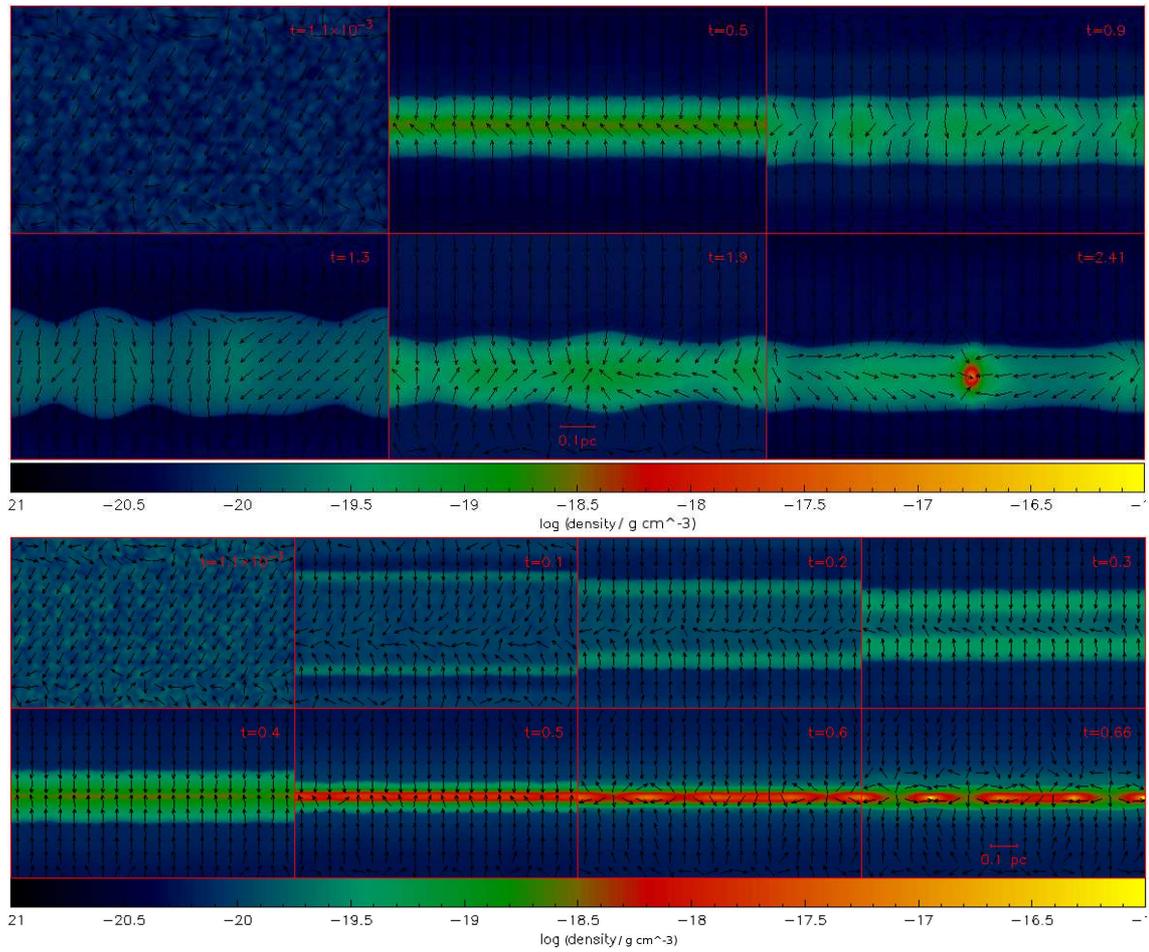

Figure A2.: As in Fig. A1 but now for respectively Cases 3 (*Upper panel*), and 4 (*Lower panel*).

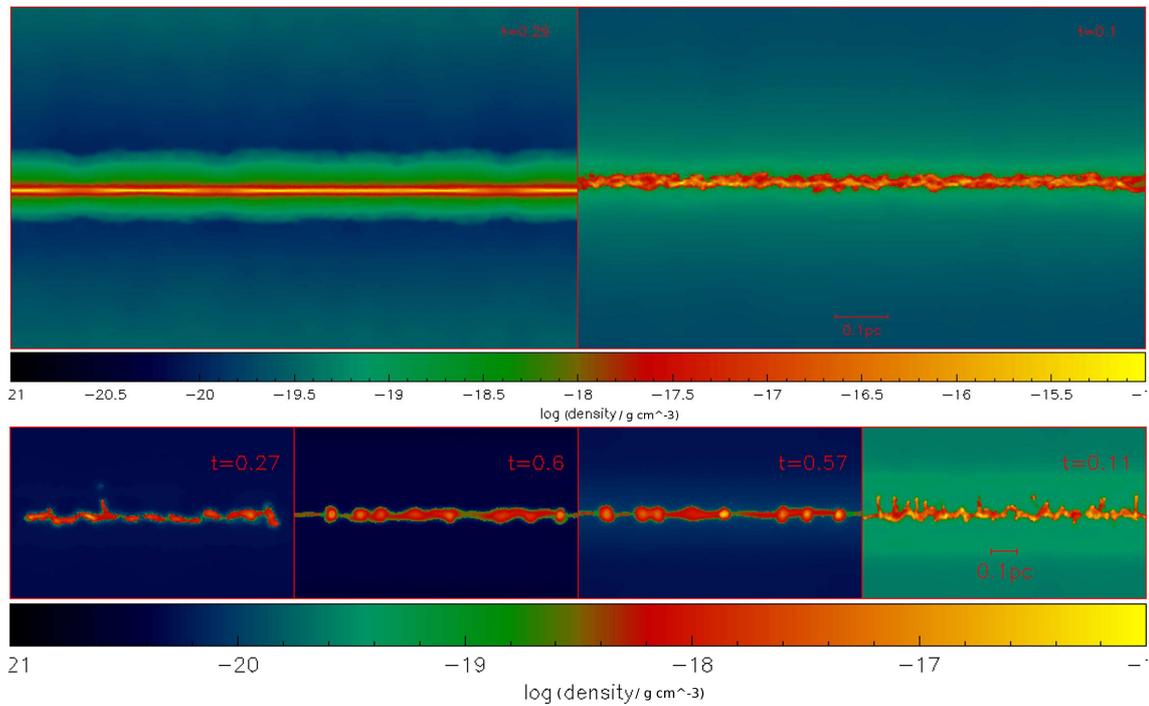

Figure A3.: As in Figs. A1 & A2 but now showing the terminal epoch of the filament in respectively Cases 5 (*Top panel left*), and 10 (*Top panel right*). Similarly the terminal epoch of the filaments in respectively Cases 6, 11, 12 & 13 are shown on the lower panel from left to right.